\documentclass[acmsmall,screen]{acmart}
\AtBeginDocument{%
  \providecommand\BibTeX{{%
    Bib\TeX}}}

\usepackage{amsmath,amsfonts}
\usepackage{bbding}
\usepackage{algorithmic}
\usepackage{graphicx}
\usepackage{textcomp}
\usepackage{xcolor}
\def\BibTeX{{\rm B\kern-.05em{\sc i\kern-.025em b}\kern-.08em
    T\kern-.1667em\lower.7ex\hbox{E}\kern-.125emX}}

\usepackage[final]{pdfpages}
\usepackage{hyperref}
\usepackage{algorithmic}
\usepackage{graphicx}
\usepackage{textcomp}
\usepackage{xcolor}
\usepackage{color}
\usepackage{url}
\usepackage{enumitem}
\usepackage{multirow}
\usepackage[ruled, vlined, linesnumbered]{algorithm2e}
\usepackage{subcaption}
\SetAlCapNameFnt{\scriptsize}
\SetAlCapFnt{\small}

\usepackage{listings}
\usepackage{verbatim}

\usepackage{xspace}

\usepackage{soul}

\sethlcolor{yellow}

\usepackage{stackengine}
\usepackage{booktabs}

\newcommand{\mypara}[1]{\noindent\textbf{#1} }

\usepackage{fbox}
\usepackage{tcolorbox}
\usepackage{fontawesome}

\usepackage{colortbl}

\usepackage{balance}
\usepackage{colortbl}
\usepackage{pgfplots}
\usepackage{microtype}
\usepgfplotslibrary{dateplot}
\pgfplotsset{compat=newest}
\usetikzlibrary{patterns}
\colorlet{shadecolor}{gray!40}
\usepackage{tikz}
\usetikzlibrary{shapes.geometric,arrows,shapes} 

\usepackage{pifont}
\usepackage{flushend}
\usepackage{ulem}

\definecolor{mygreen}{HTML}{02818a}
\definecolor{mypurple}{HTML}{8f3c8c}

\newcommand{\tool}{\textsc{MultiConf}}

\newcommand{\keepnotes}{true}

\ifthenelse{\isundefined{\keepnotes}}{
\newcommand{\mytodogreen}[1]{}
\newcommand{\civi}[1]{}
\newcommand{\yj}[1]{}
\newcommand{\yibiao}[1]{}
\newcommand{\del}[1]{}
}{
\newcommand{\mytodogreen}[1]{\textcolor{mygreen}{\ding{46}~{\sf}~#1}}
\newcommand{\yibiao}[1]{\mytodogreen{[yibiao: #1]}}
\newcommand{\yj}[1]{\textcolor{blue}{[yj: #1]}}
\newcommand{\del}[1]{\textcolor{red}{\sout{#1}}}
}

\newtcbox{\mybox}[1][darkgray]
  {on line, arc = 0pt, outer arc = 0pt,
    colback = #1!20!white, colframe = #1!50!black,
    boxsep = 0pt, left = 1pt, right = 1pt, top = 2pt, bottom = 2pt,
    boxrule = 0pt, bottomrule = 0pt, toprule = 0pt}

\begin{document}

\title{Isolating Compiler Faults via Multiple Pairs of Adversarial Compilation Configurations}

\author{Qingyang Li}
\orcid{0009-0004-5666-2158}
\email{liqingyang@smail.nju.edu.cn}

\author{Yibiao Yang}
\email{yangyibiao@nju.edu.cn}
\orcid{0000-0003-1153-2013}
\authornote{Corresponding author.}

\author{Maolin Sun}
\orcid{0000-0001-5617-2205}
\email{merlin@smail.nju.edu.cn}

\author{Jiangchang Wu}
\orcid{0000-0002-9932-7315}
\email{jiangchangwu@smail.nju.edu.cn}

\author{Qingkai Shi}
\orcid{https://orcid.org/0000-0002-8297-8998}
\email{qingkaishi@nju.edu.cn}

\author{Yuming Zhou}
\email{zhouyuming@nju.edu.cn}
\orcid{0000-0002-4645-2526}
\affiliation{%
  \institution{State Key Laboratory for Novel Software Technology, Nanjing University}
  \city{Nanjing}
  \country{China}}

\begin{abstract}
Compilers are fundamental to modern software development, making the effective identification and resolution of compiler faults essential. However, localizing these faults to specific source files remains highly challenging due to the complexity and scale of modern compiler infrastructures. 
Prior approaches that rely on passing executions of randomly generated witness test programs often incur high computational costs and produce unstable localization results. Although recent techniques that derive passing compilation configurations from failing ones have partially alleviated these limitations, substantial room for improvement remains.
In this study, we propose \tool, a novel approach that automatically isolates compiler faults by constructing multiple pairs of adversarial compilation configurations. Each adversarial compilation configuration pair consists of a failing configuration and its corresponding passing configuration, which differ in only a small number of fine-grained options. 
\tool~generates failing configurations through a lightweight construction process and derives the corresponding passing configurations by selectively disabling bug-related fine-grained options.
We then employ a Spectrum-Based Fault Localization (SBFL) formula to rank the suspiciousness of compiler source files. Each adversarial configuration pair independently produces a ranking, which is subsequently aggregated using a weighted voting scheme to derive a final suspiciousness ranking, enabling more accurate and robust fault localization. 
We evaluate \tool~on a benchmark of 60 real-world GCC compiler bugs. The results demonstrate that \tool~significantly outperforms existing compiler fault localization techniques in both effectiveness and efficiency. In particular, \tool~successfully localizes 27 out of 60 bugs at the Top-1 file level, representing improvements of 35.0\% and 28.6\% over the two state-of-the-art approaches, \textsc{Odfl} (20) and \textsc{Basic} (21), respectively.

\end{abstract}



\received{25 December 2024}
\received[revised]{07 September 2025}
\received[accepted]{19 November 2025}

\maketitle

\section{Introduction}
\label{sec:intro}
Compilers play a crucial role in software development by translating high-level programming code into executable machine code. However, the compilation phase can introduce bugs, resulting in compiler faults~\cite{issta16sun-understandbugs}. Such faults may cause incorrect program outputs, system crashes, or even security vulnerabilities. These issues pose serious challenges to software reliability and security, making both the compiler testing and fault localization crucial for trustworthy software systems.

Compiler testing identifies the presence of bugs~\cite{10.1145/1993498.1993532}, while fault localization (the focus of our work) aims to pinpoint the specific source files or components that are responsible for these bugs~\cite{chen2019compiler}.
A large body of research has focused on compiler testing (i.e., automatically finding scenarios where compilers misbehave)~\cite{emi-pldi,Yang2011,atc2025,Clozemaster,10.1145/3763094,DBLP:journals/tse/WangCCXYLDS25}. Researchers have proposed diverse approaches for compiler bug detection, as surveyed by Chen et al.~\cite{chen2020survey}. These include random program generation~\cite{Yang2011}, differential testing~\cite{Sheridan2007, Ofenbeck2016, Morisset2013,devil-asplos}, and metamorphic testing~\cite{sosp2023,Le2015}, which have significantly improved the ability to uncover subtle compiler faults. Notably, recent works have explored compiler configuration spaces to boost bug detection effectiveness, such as exploring various optimization settings~\cite{chen2022boosting}, searching two-dimensional input spaces for JIT compilers~\cite{jia2023detecting}, attribute-guided compilation space exploration~\cite{atc2025}, and auto-tuning via critical flag selection~\cite{zhu2023autotuning}. Other research has focused on detecting missed optimizations~\cite{barany2018differential} and warning defects~\cite{sun2016warning}.

Once a bug is detected, the next challenge is fault localization: determining which specific source files are responsible. This task is non-trivial, given the intricate and extensive nature of modern compilers. To illustrate the scale of the challenge, consider the GCC compiler: the average number of potentially relevant C source files per bug scenario is 380 (i.e., .c files in the top-level /trunk/gcc directory, excluding subdirectories), while the entire codebase contains more than 1,600 .c, .cc, and .h files (recursively searched in /trunk/gcc, excluding test files). This considerable scale highlights the practical difficulty of pinpointing the faulty component and motivates the need for more efficient and accurate fault localization methods.

Several approaches have been proposed for compiler fault localization. Witness test program-based methods, such as DiWi~\cite{chen2019compiler}, RecBi~\cite{chen2020enhanced}, LLM4CBI~\cite{LLM4CBI}, and ETEM~\cite{ETEM}, rely on generating passing and failing test programs to isolate faults. A failing test program is specifically designed to expose a compiler bug, whereas a passing test program is expected not to trigger the same bug under the identical compilation configuration. With these test programs, the coverage spectra can be obtained from the corresponding execution traces. Subsequently, the Spectrum-Based Fault Localization (SBFL) techniques can be applied to calculate the suspiciousness scores of code entities in compilers~\cite{abreu2009spectrum, tang2017accuracy}. 
However, as pointed out by Yang et al.~\cite{ODFL}, such approaches suffer from inherent limitations, including high computational overhead and instability due to randomness in mutation process. To address these limitations, Yang et al. proposed an alternative approach, \textsc{Odfl}~\cite{ODFL}, which constructs a single pair of adversarial compilation configurations to efficiently generate passing and failing executions for SBFL-based fault localization. 

While \textsc{Odfl} demonstrates improved accuracy, it considers only a single pair of adversarial configurations. This design choice limits its ability to fully exploit diverse coverage information, as the effectiveness of SBFL critically depends on both the quality and the quantity of the test executions~\cite{keller2017critical, tosem2013xie}. 
Motivated by ensemble learning, where aggregating multiple models typically yields more robust and accurate performance~\cite{ZHOU2002239,zhou2025ensemble}, we observe that leveraging multiple pairs of adversarial compilation configurations can provide complementary diagnostic information.
In practice, different strategies can be employed to obtain multiple pairs of adversarial compilation configurations, as different pairs possess varying capabilities for isolating irrelevant code elements. 
For instance, consider a compiler consisting of four files: A, B, C, and D, where D is the faulty file responsible for a bug. One configuration pair may isolate innocent files A and B, while another may isolate innocent files B and C. In isolation, neither pair can accurately identify the truly faulty file D. However, by aggregating the innocent files identified across these pairs, we can effectively infer that D is responsible for the bug. 
Consequently, generating multiple pairs of adversarial compilation compilations can enhance the ability to isolate compiler faults. 

\mypara{Approach.} 
In this study, we propose \tool, a novel method for isolating compiler faults by aggregating individual rankings derived from multiple pairs of adversarial compilation configurations. Our key insight is that different pairs possess varying capabilities for isolating irrelevant code responsible for compiler bugs. By aggregating the SBFL ranking results from these diverse pairs, we achieve more accurate localization results that surpass what any individual pair can provide in isolation.
We employ various strategies to generate multiple pairs of adversarial compilation configurations, ensuring that each pair maintains internal similarity. This strategic combination enables us to effectively leverage the strengths of each configuration pair, resulting in improved fault localization.

Given the significant number of source files within the compiler, our initial step is to narrow down the pool of potential faulty files using execution traces. To generate adversarial compilation configurations, we systematically disable each fine-grained option and identify bug-triggering options that convert a failing configuration into a passing one. These bug-triggering options represent potential root causes of the bug.
Next, we generate multiple pairs of adversarial compilation configurations by selecting different failing configurations. Each failing configuration produces corresponding passing configurations by selectively disabling the identified bug-triggering options.
Subsequently, we employ SBFL techniques to calculate the suspiciousness levels of compiler files for each configuration pair. This analysis enables us to rank candidate faulty compiler files based on their suspiciousness values, providing a measure of their likelihood of containing faults.
Finally, we prioritize files that consistently appear as suspicious across the majority of rankings using a weighted voting system. By aggregating these rankings, we can identify potentially faulty files.

Our evaluation of \tool~using 60 real GCC bugs demonstrates its effectiveness in localizing compiler bugs. In the Top-1, Top-5, Top-10, and Top-20 files of the ranking lists, \tool~successfully localizes 27, 40, 51, and 53 compiler bugs out of the total 60 GCC bugs, respectively.
The most significant improvement is observed in Top-1 accuracy, where \tool~achieves an 35.0\% enhancement over \textsc{Odfl}, the state-of-the-art SBFL technique (localizing 27 bugs versus \textsc{Odfl}'s 20). This advancement is particularly valuable for real-world debugging, as developers typically inspect the highest-ranked suspicious files.
Given the average of 380 source files considered per bug, these results highlight the practical utility and robustness of our multi-pair strategy.

\addvspace{5pt}

\mypara{Contributions.} 
We make the following main contributions:
\begin{itemize}[leftmargin=*]
    \item \textit{\textbf{Methodology:}}~We introduce a systematic procedure for constructing multiple pairs adversarial compilation configurations, each consisting of a failing configuration and a corresponding passing configuration with adversarial relationships. The procedure selectively disables fine-grained options within failing configurations to identify options that potentially trigger compiler bugs. 
    \item \textit{\textbf{Originality:}}~We propose a novel approach that strategically leverages multiple pairs of adversarial compilation configurations and aggregates multiple ranking results to precisely pinpoint compiler faults. To the best of our knowledge, we are the first to leverage multiple pairs of adversarial compilation configurations for compiler fault localization.
    \item \textit{\textbf{Implementation:}}~We implement our approach as a practical fault localization tool, \tool, which is powered with multiple pairs of adversarial compilation configurations. \tool~provides a lightweight yet effective solution for localizing compiler faults in practice. 
    \item \textit{\textbf{Evaluation:}}~We conduct extensive empirical evaluations of \tool~alongside existing techniques on a benchmark of 60 real-world GCC compiler bugs. The results demonstrate substantial improvements in Top-1 localization accuracy, highlighting the practical effectiveness of our approach for identifying faults in GCC.
\end{itemize}

\mypara{Organization.} 
The rest of this paper is organized as follows. 
Section~\ref{sec:background} introduces the background and presents our motivation of leveraging multiple pairs of adversarial compilation configurations.
Section~\ref{sec:approach} elaborates on our approach in detail. 
Then, our experimental setup and results are presented in Section~\ref{sec:evaluation}.
After that, we discuss the limitations, threats to the validity, and future work in Section~\ref{sec:validity}.
Finally, Section~\ref{sec:related-work} surveys related work and Section~\ref{sec:conclusion} concludes this paper.

\section{Background \& Motivation}
\label{sec:background}
In this section, we first briefly describe the SBFL technique utilized in our study. 
Then, we introduce the background of the compiler options and compiler faults. 
Finally, we summarize existing techniques in locating compiler faults, discuss their limitations, and offer a clear and concise explanation of the motivation behind our approach. 

\subsection{Spectrum-Based Fault Localization}

Spectrum-Based Fault Localization (SBFL) has received significant attention in the research community as a debugging technique to identify the root cause of software faults or failures~\cite{abreu2009spectrum, tosem2013xie}. It leverages coverage information from executed code to assess the suspiciousness of individual code statements or blocks, subsequently ranking them based on their likelihood of being faulty~\cite{abreu2009spectrum}. This approach has been extensively studied in academic literature and is recognized for its potential to guide fault localization in various scenarios.
The fundamental principle of SBFL is based on the assumption that code statements or blocks executed more frequently by failing test cases are more likely to contain faults, while those executed more frequently by passing test cases are less likely to be faulty. By ranking code elements according to their suspiciousness scores, SBFL provides developers with a prioritized list of code locations to examine, potentially reducing the time and effort required for debugging.
However, the effectiveness of SBFL heavily depends on the quality and quantity of test executions~\cite{keller2017critical, tosem2013xie}. In particular, a lack of diverse failing test cases or inadequate representation of the fault in test cases can compromise its accuracy. A previous study~\cite{keller2017critical} has demonstrated that increasing the number of failing test cases enhances the effectiveness of SBFL by providing more diverse coverage information, which is critical for accurate fault localization.
Despite its success in research settings, there is limited evidence to suggest that SBFL has been widely adopted in the software industry. Its application in practice may be constrained by challenges such as the need for comprehensive test suites and the high computational cost of analyzing complex systems. Nonetheless, its potential to assist in debugging, as shown in research studies, continues to drive interest in further refinement and application of SBFL techniques.

\subsection{Compiler Faults}

Compiler faults are errors within the compiler that can cause incorrect program execution or unexpected behavior. Notably, an empirical study of compiler bug reports revealed that optimization faults constitute the largest proportion of reported compiler issues~\cite{zhou2021empirical}. This prevalence underscores the significant impact of optimization-related bugs on software reliability and maintainability. These faults often stem from the complexity of compiler optimization algorithms and their intricate interactions with source code, making them difficult to detect and fix.
Given their frequency and the critical role of compilers in software development, there is a clear need for effective techniques to localize and address optimization faults. This motivates our study, which focuses on improving fault localization specifically for fine-grained compiler optimization options.

\subsection{Localization of Compiler Faults}

Localizing compiler faults is inherently challenging due to the complexity and scale of modern compiler infrastructures. During the code generation phase, compilers apply numerous optimizations whose intricate interactions can introduce subtle and hard-to-diagnose bugs~\cite{10.1145/1993498.1993532}.
Diagnosing and localizing optimization-related bugs is difficult because of the combinatorial explosion of possible optimization configurations, making manual debugging both time-consuming and error-prone.

To address these challenges, a variety of automated compiler fault localization techniques have been proposed. From the perspective of how passing executions are constructed to contrast with failing ones, existing approaches can be broadly classified into two categories: program mutation based localization and compilation configuration-based localization. 

\begin{itemize}[leftmargin=*]
    \item \textbf{Program Mutation-Based Localization}: These approaches generate passing executions by mutating the original bug-triggering programs to produce witness programs that no longer trigger the bug. By comparing the compiler execution traces of failing programs and their corresponding witness programs, SBFL techniques can be applied to identify suspicious fault locations. Representative techniques include DiWi~\cite{chen2019compiler}, which synthesizes witness programs using local operation mutations combined with Metropolis-Hastings sampling; RecBi~\cite{chen2020enhanced}, which employs reinforcement-learning-guided structural mutations; LLM4CBI~\cite{LLM4CBI}, which leverages large language models to generate effective witness programs; and ETEM~\cite{ETEM}, which proposes enhanced mutation strategies, simplified optimization configurations, and improved suspiciousness formulas for file-level localization. 
    \item \textbf{Compilation Configuration-Based Localization}: Instead of modifying test programs, this line of work generates passing executions by manipulating compiler optimization configurations. \textsc{Odfl}~\cite{ODFL}, for example, systematically disables fine-grained optimization options in a failing compilation configuration to construct adversarial passing--failing configuration pairs for the same test program. By contrasting compiler behaviors under these configurations, fault localization can be effectively performed. 
\end{itemize}

Program mutation-based approaches rely on the availability of both failing and passing test programs. While bug-triggering programs are designed to expose specific compiler faults, witness programs are crafted to avoid triggering the bug while remaining semantically similar. The resulting contrast between passing and failing executions enables SBFL techniques to identify potential faulty components within the compiler. In contrast, compilation configuration-based approaches operate on a fixed test program and instead explore the optimization configuration space. For a given program, some configurations lead to failing compilations, whereas others succeed. By systematically identifying such adversarial configuration pairs, these approaches similarly obtain sufficient execution diversity to support SBFL-based fault localization, without incurring the substantial cost associated with program-level mutations.

\begin{figure}
    \centering
    \begin{subfigure}{0.32\textwidth} %
        \centering
        \begin{lstlisting}[basicstyle=\scriptsize\ttfamily]
        int a;
        int b;
        int *c = &a;
        unsigned short d;
        
        int main () {
          unsigned int e = a;
          *c = 1;
          if (!b) {
              d = e;
              *c = d | e;
            }
          if (a != 0) {
            __builtin_abort (); 
          }
          return 0;
        }
        \end{lstlisting}
    
        \lstset{escapeinside={<@}{@>}}
        \caption{Bug-triggering test program g.c}\label{fig:example-gcc-a}
    \end{subfigure}
    \hspace{0.2em}  %
    \begin{subfigure}{0.46\textwidth}
        \centering
        \begin{lstlisting}[basicstyle=\scriptsize\ttfamily,showlines=True]
        $ (*@\mybox[green]{gcc -O1 g.c}@*) && ./a.out
        $ (*@\mybox[red]{gcc -Os g.c}@*) && ./a.out
        (*@\mybox[red]{\textbf{Aborted (core dumped)}}@*)
        $ (*@\mybox[green]{gcc -Os \textcolor{blue}{\textbf{-fno-expensive-optimizations}} g.c}@*) & ./a.out
        \end{lstlisting}
        \caption{Compile configurations}\label{fig:example-gcc-b}
    \end{subfigure}
    
    \caption{
    GCC bug \href{https://gcc.gnu.org/bugzilla/show_bug.cgi?id=61517}{\#61517}.
    When the test program is compiled at the ``-O1'' optimization level, it executed as expected and did not produce any output. 
    While at the ``-Os'' optimization level, the program aborted and generated a core dump.
    }
    \label{fig:example-gcc}
\end{figure}

\subsection{Discussion of Existing Techniques}

Existing techniques have demonstrated effectiveness in locating compiler bugs. However, as pointed out by Yang et al.~\cite{ODFL}, witness test program-based approaches, such as DiWi~\cite{chen2019compiler}, RecBi~\cite{chen2020enhanced}, LLM4CBI~\cite{LLM4CBI}, and ETEM~\cite{ETEM}, can be time-consuming and unstable due to the inherent randomness in the mutation process.

The adversarial configuration-based approach \textsc{Odfl}~\cite{ODFL} has demonstrated that the optimization processes associated with some particular options may serve as the root cause of bugs and passing configurations can be generated by disabling these bug-triggering options. As shown in Figure~\ref{fig:example-gcc}, disabling specific fine-grained optimization options in GCC can effectively hide compiler optimization bugs that are initially exposed by the bug-triggering test program.

However, the compilation configuration space is extensive, with a size of $2^{n}$, where $n$ represents the number of fine-grained compilation options that are enabled by default at the bug-triggering level. In most cases, $n$ exceeds 100 (as shown in Table~\ref{tbl:optnum}), making it challenging to select appropriate configurations for analysis. While some configurations may accurately identify faulty files, others may not, resulting in suboptimal performance of the generated configurations. 
\textsc{Odfl} is limited by its reliance on a single pair of adversarial compilation configurations, which may fail to capture sufficient coverage diversity for optimal localization performance. Although ensuring similarity among these configurations is essential for distinguishing innocent files from suspicious ones, this focus also restricts the exploration of a broader range of configurations, potentially compromising the accuracy of bug localization results.

On one hand, achieving comprehensive coverage requires expanding the range of adversarial compilation configurations. On the other hand, maintaining similarity among configurations is crucial for more targeted analysis, as similar execution traces enable a more precise assessment of code coverage changes, which helps in distinguishing suspicious files.
Initially, these requirements may seem contradictory. To address this challenge, we draw inspiration from ensemble learning, which combines the predictions of multiple base learners to create a more powerful and robust model. We generate multiple pairs of adversarial compilation configurations that maintain similarity within each pair while differing among pairs. By aggregating the localization results obtained from these pairs, we achieve a more robust and reliable outcome for fault localization.

\begin{table}[th]
    \renewcommand{\arraystretch}{1.2}
    \setlength{\tabcolsep}{10pt}
    \vspace{-1.0em}
    \caption{Statistics of the fine-grained optimization options enabled by default at the lowest bug-triggering optimization level for the bugs in the benchmark.}
    \vspace{1.0em}
    \label{tbl:optnum}
    \centering
    \footnotesize
    \begin{tabular}{ccccc}
        \toprule
        \textbf{Bug-triggering level} & \textbf{Minimum} & \textbf{Maximum} & \textbf{Median} & \textbf{Mean} \\
        \midrule
         Lowest  & 78  & 141  & 119 & 116  \\
       \bottomrule
    \end{tabular}
    \vspace{-1.0em}
\end{table}

\section{Approach}
\label{sec:approach}
Based on the insights obtained in Section~\ref{sec:background}, we introduce \tool, a novel approach for localizing compiler faults, which leverages multiple pairs of adversarial compilation configurations. In this section, we present an overview of~\tool, provide a comprehensive explanation of our approach, and describe the SBFL formula employed by \tool.

\begin{figure}
    \includegraphics[width=0.9\textwidth]{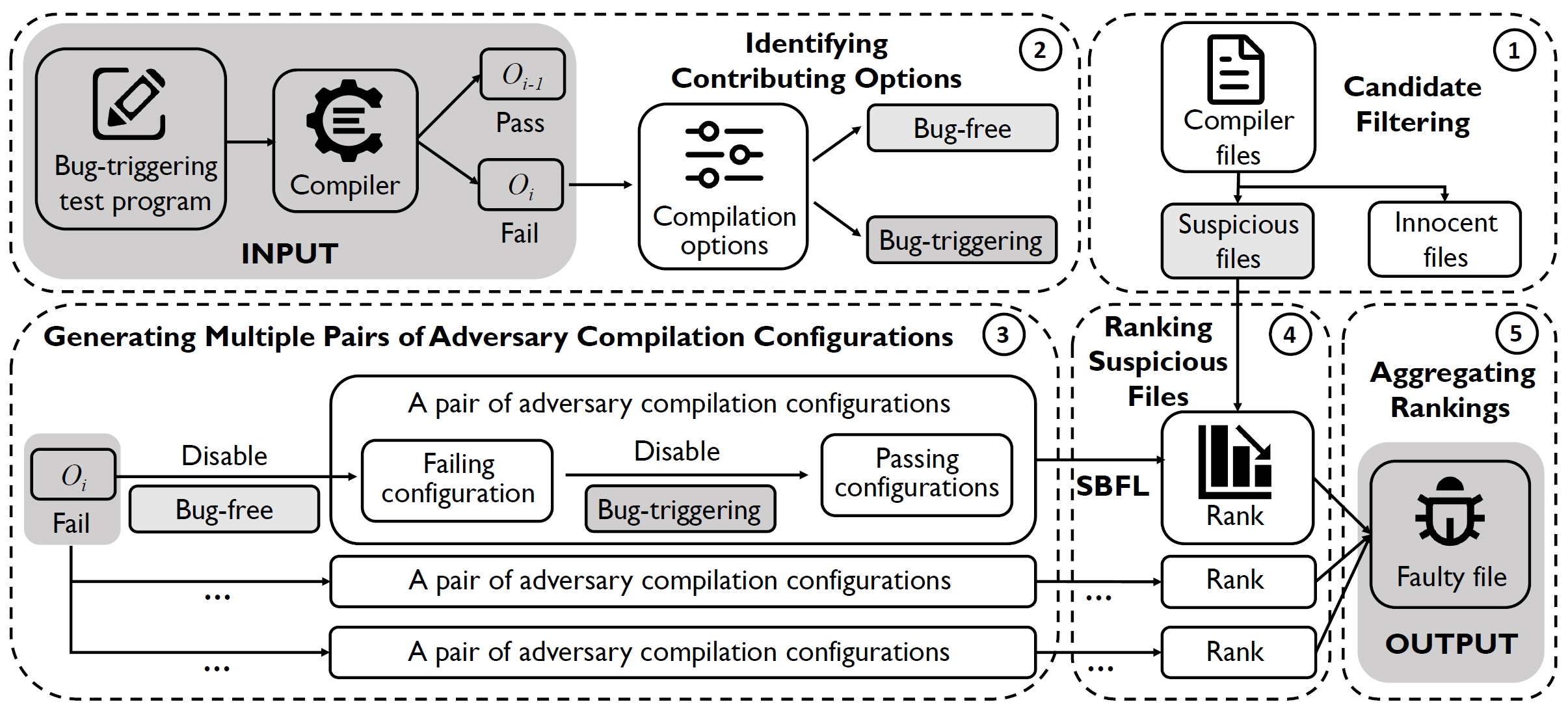}
    \caption{An overview of~\tool.}
    \label{fig:overview}
\end{figure}

\subsection{Overview}
\label{sec:approach:overview}

\newcommand\mycommfont[1]{\footnotesize\ttfamily\textcolor{blue}{#1}}
\SetCommentSty{mycommfont}

\setlength{\textfloatsep}{0.5cm}

\begin{algorithm}[htp]
\SetKwFunction{FnCollectCoverage}{CollectCoverage}
\SetKwFunction{FnGetSuspiciousFiles}{GetSuspiciousFiles}
\SetKwFunction{FnGetEnabledFineGrainedOptions}{GetEnabledFineGrainedOptions}
\SetKwFunction{FnGenAdvConfs}{GenAdvConfs}
\SetKwFunction{FnGetHalfElement}{GetHalfElement}
\SetKwFunction{FnDisableOptions}{DisableOptions}
\SetKwFunction{FnEnableOptions}{EnableOptions}
\SetKwFunction{FnGetResult}{GetResult}
\SetKwFunction{FnSBFL}{SBFL}
\SetKwFunction{FnRankAggr}{RankAggr}
\SetKwProg{Fn}{Function}{}{end}
\SetKwProg{Fp}{Procedure}{}{end}

\caption{\small Compiler Faults Isolation with \tool}
\label{ago:overall algo}
\DontPrintSemicolon
\small

\KwIn{
    $\mathit{O_\mathit{fail}}$: The lowest optimization level reproducing the bug; \\
    $\mathit{O_\mathit{pass}}$: The highest optimization level concealing the bug.
}
\KwOut{
    $\mathit{Rank}$: The aggregated rank of suspicious compiler files.
}

\BlankLine 

\tcc{Step1: Candidate Filtering}
$\mathit{FailCovSet} \leftarrow$ \FnCollectCoverage($\mathit{O_\mathit{fail}}$) \;
$\mathit{PassCovSet} \leftarrow$ \FnCollectCoverage($\mathit{O_\mathit{pass}}$) \;
$\mathit{SuspiciousCovSet} \leftarrow \mathit{FailCovSet} - \mathit{PassCovSet}$ \;
$\mathit{SuspiciousFiles} \leftarrow$ \FnGetSuspiciousFiles($\mathit{SuspiciousCovSet}$) \;
    
\BlankLine 

\tcc{Step2: Identifying Contributing Options}
$\mathit{PassResult} \leftarrow$ \FnGetResult($\mathit{O_\mathit{pass}}$) \;
$F \leftarrow$ \FnGetEnabledFineGrainedOptions($\mathit{O_\mathit{fail}}$) \;
$F_\mathit{bt} \leftarrow \varnothing$ \tcc*[l]{$F_\mathit{bt}$: the set of bug-triggering options}
\ForEach{$f \in F$}{
    $\mathit{conf} \leftarrow$ \FnDisableOptions($O_\mathit{fail}, \{f\}$) \;
    \If{\FnGetResult($\mathit{conf}$) $== \mathit{PassResult}$}{
        $F_\mathit{bt} \leftarrow F_\mathit{bt} \cup \{f\}$ \;
    }
}
$F_\mathit{bf} \leftarrow F - F_\mathit{bt}$\tcc*[l]{$F_\mathit{bf}$: the set of bug-free options}

\BlankLine 

\tcc{Step3: Generating Adversarial Configurations}
$\mathit{AdvConf}_1 \leftarrow$ \FnGenAdvConfs($O_\mathit{fail}, F_\mathit{bf}, F_\mathit{bt}$) \;
$\mathit{AdvConf}_2 \leftarrow$ \FnGenAdvConfs($\mathit{O_\mathit{fail}}$, \FnGetHalfElement($F_\mathit{bf}$), $F_\mathit{bt}$) \;
$\mathit{AdvConf}_3 \leftarrow$ \FnGenAdvConfs($O_\mathit{fail}, \varnothing, F_\mathit{bt}$) \;
    
\BlankLine   

\tcc{Step4: Ranking Suspicious Files}
$\mathit{Rank}_{1} \leftarrow$ \FnSBFL($\mathit{AdvConf}_{1}, \mathit{SuspiciousFiles}$) \;
$\mathit{Rank}_{2} \leftarrow$ \FnSBFL($\mathit{AdvConf}_{2}, \mathit{SuspiciousFiles}$) \;
$\mathit{Rank}_{3} \leftarrow$ \FnSBFL($\mathit{AdvConf}_{3}, \mathit{SuspiciousFiles}$) \;
    
\BlankLine

\tcc{Step5: Aggregating Rankings}
$\mathit{Rank} \leftarrow$ \FnRankAggr($\mathit{Rank}_{1}, \mathit{Rank}_{2}, \mathit{Rank}_{3}$) \;
    
\BlankLine  
\Return{$\mathit{Rank}$} \;

\BlankLine
\BlankLine
\Fn{GenAdvConfs($O_\mathit{fail}, F_\mathit{disable}, F_\mathit{bt}$)}{
    \tcc{$F_\mathit{disable}$: the selected bug-free options from $F_\mathit{bf}$}
    $\mathit{FailConf} \leftarrow$ \FnDisableOptions($O_\mathit{fail}, F_\mathit{disable}$) \;
    $\mathit{PassConfs} \leftarrow \varnothing$ \;
    \ForEach{$F_\mathit{bt} \in F_\mathit{bt}$}{
        $\mathit{conf} \leftarrow$ \FnDisableOptions($\mathit{FailConf}, \{F_\mathit{bt}\}$) \;
        $\mathit{PassConfs} \leftarrow \mathit{PassConfs} \cup$ \{$\mathit{conf}$\} \;
    }
    \tcc{$\mathit{FailConf}$: the failing configuration}
    \tcc{$\mathit{PassConfs}$: a set of passing configurations created by disabling each bug-triggering option in $\mathit{FailConf}$}
    \Return{$(\mathit{FailConf}, \mathit{PassConfs})$} \;
}

\end{algorithm}

We present \tool, a novel five-phase methodology for compiler fault isolation that significantly advances the \textsc{Odfl} framework in both effectiveness and efficiency. Figure~\ref{fig:overview} provides an illustration of our approach.

Although inspired by \textsc{Odfl} and conceptually a multi-pair generalization, \tool~introduces several essential architectural and algorithmic innovations. Specifically, \tool~evolves the single-pair analysis of \textsc{Odfl} into a robust multi-pair framework, incorporates novel candidate filtering and weighted aggregation mechanisms, and achieves substantial practical efficiency improvements. These enhancements yield marked gains in fault localization accuracy and robustness, particularly for challenging compiler bugs (see Table~\ref{tbl:comparison} and Table~\ref{tbl:efficiency}).

\tool~takes four key inputs: (1) the compiler, (2) a bug-triggering test program, (3) the lowest optimization level that triggers the bug, and (4) the highest optimization level where the bug does not occur. Leveraging these inputs, \tool~systematically constructs adversarial configuration pairs to guide fault localization. The main output is a ranked list of suspicious compiler source files, along with the identification of optimization options critical to bug manifestation\textemdash providing actionable diagnostic insights for developers.
The core workflow, outlined in Algorithm~\ref{ago:overall algo}, proceeds through the following phases:
\begin{enumerate}[leftmargin=*]
\item \textbf{Candidate Filtering:} Unlike \textsc{Odfl}, which considers all compiler files as possible fault locations, \tool~applies a two-stage, coverage-driven filtering to greatly reduce the search space (Lines 1–4). This step narrows the candidate set by 66\% on average (from 380 to 129 files), improving both efficiency and precision.
\item \textbf{Identifying Contributing Options:} Like \textsc{Odfl}, \tool~identifies fine-grained optimization options directly involved in triggering the bug (Lines 5–12).
\item \textbf{Generating Multiple Adversarial Configurations:} In contrast to \textsc{Odfl}'s single-pair approach, \tool~systematically constructs multiple adversarial configuration pairs from distinct failing configurations (Lines 13–15). Each passing configuration in the pair is created by independently disabling a bug-triggering option. This strategy captures a wider spectrum of failure-inducing conditions. In our experiments, we generate three such pairs per bug instance.
\item \textbf{Ranking Suspicious Files:} Each adversarial pair undergoes spectrum-based fault localization (SBFL) analysis, as in \textsc{Odfl}, producing multiple independent suspiciousness rankings (Lines 16–18). This multi-perspective analysis increases localization robustness, as different pairs may highlight different aspects of the underlying fault.
\item \textbf{Aggregating Rankings:} \tool~introduces a novel weighted voting scheme to aggregate rankings from all adversarial pairs (Line 19). Files consistently ranked as highly suspicious across pairs receive higher priority, yielding a more reliable and discriminative identification of faulty code regions.
\end{enumerate}

\tool~achieves substantial advancement over prior work in three key areas:

\begin{itemize}[leftmargin=*]
    \item \textit{Architectural Generalization}: \tool~systematically integrates diagnostic evidence from multiple configuration pairs, addressing the instability and coverage limitations inherent to single-pair approaches like \textsc{Odfl}. This architecture not only improves accuracy in fault localization, but also enables new forms of analysis that are not possible with single-pair methods.
    \item \textit{Algorithmic Contributions}: Beyond the multi-pair paradigm, \tool~introduces two novel algorithmic components: candidate filtering and weighted aggregation. These innovations optimize both the efficiency and the discriminative power of the fault localization process, and can be generalized to other configuration-based or program-based localization frameworks.
    \item \textit{Practical Improvement}: \tool~features an improved adversarial configuration generation strategy, yielding a 93\% reduction in generation time compared to \textsc{Odfl} (see Table~\ref{tbl:efficiency}). This makes the approach more practical and scalable for real-world use.
\end{itemize}

Overall, \tool~offers clear advantages over \textsc{Odfl} by providing (1) robust performance across diverse configuration pairs, and (2) effective exploitation of varied coverage patterns from multiple configurations. Experimental results demonstrate that \tool~maintains \textsc{Odfl}'s computational efficiency while delivering substantially higher localization accuracy, particularly for complex optimization bugs where traditional single-pair approaches are less effective.

\subsection{Key Steps}

To isolate compiler faults, \tool~employs a five-step process that involves multiple pairs of adversarial compilation configurations. We provide a detailed outline of each key step below.

\subsubsection{Candidate Filtering}

We begin by defining the candidate scope and ``clean'' files for each bug. Candidates include all C source files in the main GCC directory (i.e., *.c files under gcc), averaging 380 files per bug. The faulty file(s) are those modified by the developer when fixing the bug; ``clean'' files are all remaining files.

We then reduce the candidate set via two filtering steps (Lines 1-4), summarized in Table~\ref{tbl:susfilenum}:

\begin{itemize}[leftmargin=*]
    \item \textbf{Coverage presence filter.} We remove files that are not executed when compiling the bug-triggering program under the bug-triggering configuration. The remaining ``Files Covered'' retain only executed files relevant to the compilation.
    \item \textbf{Differential spectrum filter.} We compare spectra between the failing and passing configurations and remove files whose coverage is identical across the two, as they provide no discriminative signal for SBFL. The remaining ``Files with Diff Cov'' are those with differing spectra across the initial adversarial pair.
\end{itemize}

Across the benchmark, these two steps reduce the average candidate set from 380 to 129 files, substantially shrinking the search space without excluding any true faulty files in our dataset. This improves both the efficiency and the robustness of subsequent SBFL and aggregation.

\begin{table}[ht]
    \renewcommand{\arraystretch}{1.0}
    \setlength{\tabcolsep}{3pt}
    \caption{Statistic of the decreased number of suspicious files after filtering for the bugs in the benchmark.}
    \vspace{1.0em}
    \label{tbl:susfilenum}
    \centering
    \footnotesize
    \begin{tabular}{lcccccc}
        \toprule
        \textbf{} & \textbf{Average} & $\Downarrow$ & \textbf{Maximum} & $\Downarrow$ & \textbf{Minimum} & $\Downarrow$ \\
        \midrule
        Compiler Files & 380 & — & 433 & — & 359 & — \\
        Files Covered & 304 & 20.00 & 345 & 20.32 & 169 & 52.92 \\
        Files with Diff Cov & \textbf{129} & 66.05 & \textbf{197} & 54.50 & \textbf{52} & 85.52  \\
       \bottomrule
    \end{tabular}
    \vspace{-1.0em}
    \caption*{\footnotesize Columns ``$\Downarrow$'' refers to the decreased rate after filtering.}
\end{table}

\subsubsection{Identifying Contributing Options}
The second step of our approach aims to identify the specific fine-grained options that directly contribute to the occurrence of the bug. This step is motivated by the observation that most bugs can be concealed when lower levels of coarse-grained optimization or no optimization are applied. Therefore, it is reasonable to assume that the occurrence of optimization bugs is associated with certain fine-grained options.

To facilitate further analysis and discussion, we introduce the following definitions:
\begin{itemize}[leftmargin=*]
    \item \textbf{Bug-Triggering Option:} A fine-grained option is categorized as bug-triggering if disabling it at the bug-triggering optimization level conceals the bug.
    \item \textbf{Bug-Free Option:} A fine-grained option is categorized as bug-free if disabling it at the bug-triggering optimization level still triggers the bug.
\end{itemize}

To effectively narrow down the search space, we begin by selecting the lowest coarse-grained optimization level that reproduces the bug. For instance, if ``-O0'' conceals the bug while both ``-O1'' and ``-O2'' reproduce it, we select ``-O1''. This approach minimizes the number of enabled fine-grained optimizations.
We then systematically disable each fine-grained optimization option individually (Line 9). By observing the program's behavior and output, we can identify the bug-triggering options. If disabling a specific option successfully conceals the bug—meaning the program's behavior and output align with expectations—we categorize that option as a bug-triggering option (Lines 10-11). The remaining options are classified as bug-free options (Line 12). This categorization process allows us to pinpoint the fine-grained optimizations directly responsible for triggering the bug and facilitates further analysis.

\subsubsection{Generating Adversarial Configurations}
In the third step of our approach, our objective is to generate multiple pairs of adversarial compilation configurations using the bug-triggering and bug-free options identified in the previous step. This process allows us to examine the specific contributions of these optimizations to the bug from different perspectives.

For a specific test program, different compiler setups can yield varying results: some configurations (failing configurations) trigger the bug, while others (passing configurations) do not. These configurations are considered adversarial because they exhibit distinct behaviors with respect to the bug. Our goal is to construct adversarial configurations that differ by only a few options.

In our approach, the \texttt{GenAdvConfs} procedure produces an $\mathit{AdvConf}$ pair, comprising a failing configuration and a set of passing configurations. Formally, each pair is denoted as $(\mathit{FailConf}, \mathit{PassConfs})$, where $\mathit{FailConf}$ is a configuration that manifests the bug, and $\mathit{PassConfs}$ is a collection of configurations that do not exhibit the bug. Each passing configuration in $\mathit{PassConfs}$ is derived by disabling exactly one bug-triggering option present in the failing configuration, ensuring that it differs from $\mathit{FailConf}$ by only a single option (Lines 24–26).
Consequently, every $(\mathit{FailConf}, \mathit{PassConfs})$ pair forms an adversarial relationship suitable for SBFL analysis.

For this study, we select as our initial failing configuration the lowest optimization level at which the bug can be reproduced. We then generate three distinct types of failing configurations by disabling different subsets of bug-free options, as follows:
\begin{itemize}[leftmargin=*]
\item \textbf{$\mathit{FailConf}_{1}$}: disables all bug-free options (Line 13).
\item \textbf{$\mathit{FailConf}_{2}$}: disables half of the bug-free options (Line 14).
\item \textbf{$\mathit{FailConf}_{3}$}: disables none of the bug-free options (Line 15).
\end{itemize}

By constructing these failing configurations, \texttt{GenAdvConfs} yields three corresponding adversarial pairs, $(\mathit{FailConf}, \mathit{PassConfs})$, denoted as $\mathit{AdvConf}_1$, $\mathit{AdvConf}_2$, and $\mathit{AdvConf}_3$. In each case, the passing configurations in $\mathit{PassConfs}$ are adversarial with respect to their associated $\mathit{FailConf}$.
This systematic exploration enables us to thoroughly investigate the influence of bug-triggering options and to gain deeper insights into their role in exposing the bug.

While our algorithm effectively identifies directly causative bug-triggering options, there are cases in which no such options can be isolated. This situation arises either (1) when the bug results from inherent behavior at a specific optimization level, rather than from individual fine-grained options, or (2) when the bug emerges only due to complex interactions among multiple options.
To address these scenarios, we additionally generate adversarial configuration pairs at the optimization level granularity. In these cases, each pair $(\mathit{AdvConf})$ is of the form $(\mathit{O_\mathit{fail}}, \mathit{O_\mathit{pass}})$, where $\mathit{O_\mathit{fail}}$ denotes the bug-triggering optimization level and $\mathit{O_\mathit{pass}}$ is a lower, non-bug-triggering optimization level.

\subsubsection{Ranking Suspicious Files}
In the fourth step of our approach, we employ the Spectrum-Based Fault Localization (SBFL) technique to localize faults within each pair of collected adversarial compilation configurations. SBFL is based on the assumption that code statements executed frequently in different passing cases are less likely to be the cause of the fault, while specific code statements executed due to the enabling of certain bug-triggering options are more likely to be the source of the issue.
Following the approach of previous studies~\cite{chen2019compiler, chen2020enhanced, ODFL}, our tool, \tool, utilizes the Ochiai formula~\cite{abreu2007accuracy} to calculate the suspiciousness value for each statement $s$ that is executed or not executed within the configurations. The Ochiai formula, defined in Formula~\ref{formula:ochia}, takes into account the number of bug-triggering cases ($ef_{s}$) that execute the statement $s$, the number of bug-concealing cases ($ep_{s}$) that execute the statement $s$, and the number of bug-triggering configurations ($nf_{s}$) that do not execute the statement $s$.

\begin{equation}
  suspicious(s) = \frac{ef_{s}}{\sqrt{(ef_{s}+nf_{s})(ef_{s}+ep_{s})} }\label{formula:ochia}
\end{equation}

\noindent In this study, since all passing configurations are paired with the same failing configuration that triggers the bug, and we are specifically interested in the statements executed by the failing case, we can simplify the formula as follows:

\begin{equation}
  suspicious(s) = \frac{1}{\sqrt{1+ep_{s}}}\label{formula:ochia_new}
\end{equation}

\noindent After computing the suspiciousness values for each affected statement in the suspicious compiler files identified in the first step, \tool~proceeds to calculate the overall suspiciousness value for each of these files. This is achieved by aggregating the suspiciousness values of all statements within the file, using the aggregation formula established in previous studies~\cite{chen2019compiler, chen2020enhanced, ODFL} as shown in Formula~\ref{formula:aggregate}. Here, $n_{f}$ denotes the count of statements covered by the failing configuration in the compiler file $f$.

\begin{equation}
  suspicious(f) = \frac{\sum_{i=1}^{n_{f}}suspicious(s_i)}{n_{f}} \label{formula:aggregate}
\end{equation}

\noindent Finally, the compiler files are ranked based on their suspiciousness values, with higher-ranked files indicating a greater probability of containing bugs. 

\subsubsection{Aggregating Rankings}
The final step of our approach involves aggregating the three individual ranks obtained from the previous step to produce a consolidated ranking of the compiler files. To achieve a more robust and reliable evaluation of the suspiciousness of these files, we prioritize those that consistently appear as suspicious across the majority of individual rankings, using a weighted voting system.

The weighting of the individual rankings is a crucial component of this aggregation process. In the context of compiler fault localization, the Top-N metric is a widely recognized evaluation criterion, where files ranked higher are considered more impactful and deserving of a development team's limited time and resources. These higher-ranked files typically become the focus of further investigation and debugging efforts, as they are more likely to reveal the root causes of the observed compiler issues.

We recognize that the original suspicious scores used to generate individual rankings may vary or exhibit inconsistencies across different ranking methods. This is why we utilize multiple pairs of adversarial compilation configurations to produce varied ranks. Such discrepancies can stem from the unique characteristics and assumptions of each ranking approach, as well as the inherent complexity of the compiler codebase. To address these potential differences, we base the voting power of each file on its rank value rather than solely on the raw suspicious scores.

Specifically, we assign the following voting powers to files based on their rank positions:
\begin{itemize}[leftmargin=*]
    \item Files ranked at the very top (e.g., rank 1) receive the highest voting power, such as 5 points.
    \item Files within the top-5 range receive 4 points.
    \item Files in the top-10 range receive 3 points.
    \item Files in the top-20 range receive 2 points.
    \item For files ranked outside the top-20, we use the multiplicative inverse of their rank value as the voting power (e.g., a file ranked 50th would have a voting power of \( \frac{1}{50} = 0.02 \)).
\end{itemize}

After determining the individual voting powers, we sum the three voting scores for each file to arrive at the final aggregated ranking. This rank-based weighting scheme ensures that files consistently ranked at the top positions, regardless of their specific suspicious score values, have the most influence in the final aggregated ranking.

\mypara{Case Study: GCC Bug \#70138.}
Consider GCC bug \#70138, involving two files of interest: \texttt{tree-vect-slp.c} and \texttt{tree-vect-loop-manip.c}, each evaluated across three adversarial configuration pairs.

\begin{itemize}[leftmargin=*]
    \item \texttt{tree-vect-slp.c}: ranks $1$, $1$, $67$ (scores $0.707$, $0.707$, $0.411$).
    \item \texttt{tree-vect-loop-manip.c}: ranks $4$, $3$, $4$ (scores $0.655$, $0.647$, $0.701$).
\end{itemize}

\noindent Computing aggregate scores:
\begin{align*}
    V_{\texttt{tree-vect-slp.c}} &= 5\,(\text{rank }1) + 5\,(\text{rank }1) + 1/67\,(\text{rank }67) \approx 10.015 \\
    V_{\texttt{tree-vect-loop-manip.c}} &= 4\,(\text{rank }4) + 4\,(\text{rank }3) + 4\,(\text{rank }4) = 12
\end{align*}

\noindent Despite \texttt{tree-vect-slp.c} attaining first place in two analyses, its aggregate score is reduced due to the outlier 67th rank. In contrast, \texttt{tree-vect-loop-manip.c} achieves a higher aggregate due to its stable top-5 performance.

\mypara{Analysis.}
Manual inspection confirms the true bug resides in \texttt{tree-vect-loop-manip.c}, not in \texttt{tree-vect-slp.c}. The latter's sporadically high ranks are attributed to coincidental coverage, illustrating the danger of over-reliance on single-configuration results.

\noindent Our aggregation scheme offers three key advantages:
\begin{enumerate}[leftmargin=*]
    \item \textbf{Outlier mitigation:} Files with highly inconsistent ranks are downweighted.
    \item \textbf{Stability reward:} Files consistently appearing in top positions accumulate higher scores.
    \item \textbf{Robustness:} The final ranking reflects consensus across diverse analyses, reducing the impact of configuration-specific artifacts.
\end{enumerate}

\noindent In summary, rank-based aggregation enables more reliable identification of faulty files by amplifying consistent signals and attenuating noise, thus supporting more effective fault localization in complex compiler codebases.

\section{Evaluation}
\label{sec:evaluation}
\newcommand{\cis}{C}
\newcommand{\nis}{\overline{C}}
In this study, we address the following research questions:

\begin{itemize}[leftmargin=*]
\item \textbf{RQ1}: How effective is \tool~in locating compiler faults?
\item \textbf{RQ2}: How efficient is \tool~in locating compiler faults?
\item \textbf{RQ3}: What is the impact of candidate filtering on the effectiveness of \tool?
\item \textbf{RQ4}: How does the number of adversarial compilation configuration pairs influence the performance of \tool?
\item \textbf{RQ5}: How do different SBFL formulae affect the performance of \tool?
\end{itemize}

\textit{RQ1} evaluates the effectiveness of \tool~in isolating compiler faults compared to existing techniques. Through controlled experiments on 60 real GCC bugs, we quantitatively assess \tool's accuracy in Top-N fault localization metrics (N=1,5,10,20) and compare with state-of-the-art approaches (\textsc{Odfl}, DiWi, RecBi, LLM4CBI, ETEM). This investigation aims to demonstrate \tool's capability to precisely identify faulty files in compiler.

\textit{RQ2} focuses on evaluating the efficiency of \tool~in locating compiler faults. Specifically, we aim to investigate the computational resources required, particularly the execution time needed to generate multiple pairs of compilation configurations for fault localization. By assessing the efficiency of \tool, we aim to determine whether it can provide time-efficient localization of compiler faults.

\textit{RQ3} examines the impact of candidate filtering on localization accuracy. Through ablation studies, we analyze how the two-stage filtering - first removing non-executed files, then eliminating files with identical coverage spectra - reduces the search space from 380 files to 129 files (66\% reduction) while preserving true faulty files.

\textit{RQ4} explores the impact of the number of adversarial configuration pairs on localization performance. We systematically vary the pair count (1, 3, 5, 10) while measuring: (1) Top-N accuracy, (2) ranking stability (MFR/MAR), and (3) computational cost. This reveals the optimal balance between diversity benefits and resource overhead.

\textit{RQ5} assesses \tool's robustness across six SBFL formulae (Ochiai, Tarantula, DStar, etc.). Through controlled experiments, we measure the variance in Top-N accuracy, confirming that \tool's performance is stable regardless of the underlying SBFL formula choice.

\subsection{Experimental Setup}

\subsubsection{Benchmark}

Our research focuses on GCC, a widely-used open-source C compiler~\cite{gcc-}, which has become a standard in both industry and academia due to its extensive application across diverse domains~\cite{glek2010optimizing, fursin2008milepost, caceres2017automatic, berlin2004high}.
To evaluate the effectiveness of \tool, we employ a benchmark comprising 60 real-world GCC bugs, consistent with prior studies. Each bug in this benchmark represents a confirmed issue that has been fixed by developers. The benchmark, originally released by RecBi~\cite{chen2020enhanced} and subsequently adopted by \textsc{Odfl}~\cite{ODFL}, LLM4CBI~\cite{LLM4CBI}, and ETEM~\cite{ETEM}, enables reliable and fair comparisons with prior work.

For each of the 60 GCC bugs, we collected essential information from the corresponding bug reports: the lowest optimization level that triggers the bug, the minimal test program that reproduces it, the specific GCC version containing the bug, and the compiler source files responsible for the faulty behavior. This curated dataset serves as the ground truth for evaluation. We assess \tool’s localization ability by querying the ordinal position of the true faulty files in \tool’s ranking: higher ranks (smaller numbers) indicate better localization.

We intentionally limit our evaluation to GCC rather than including LLVM to ensure a meaningful and challenging assessment. LLVM’s strictly ordered pass pipeline enables efficient binary search to locate bug-triggering passes; the identified pass typically aligns with the best failing configuration for fault localization, making multiple adversarial configuration pairs less impactful. In contrast, GCC exposes a richer, more challenging space: its many fine-grained, unordered options make it difficult to identify strong adversarial configurations. This setting better stresses and differentiates approaches like \tool.

We summarize the bug types in our benchmark in Table~\ref{tbl:bugtype}. Among the 60 bugs, 55 are wrong-code bugs and 5 are ICE on valid code.

\begin{table}[ht]
\vspace{-1.0em}
    \renewcommand{\arraystretch}{1.0}
    \setlength{\tabcolsep}{10pt}
    \centering
    \caption{Distribution of Bug Types in the Benchmark}
    \vspace{1.0em}
    \label{tbl:bugtype}
    \footnotesize
    \begin{tabular}{l r}
        \toprule
        \textbf{Bug Type} & \textbf{Count} \\
        \midrule
        Wrong code & 55 \\
        ICE on valid code & 5 \\
        \midrule
        Total & 60 \\
        \bottomrule
    \end{tabular}
\end{table}

\subsubsection{Baselines}

To evaluate the effectiveness of \tool~in isolating compiler faults, we include representative compilation configuration-based and program mutation-based localization approaches as the baselines:
\begin{itemize}[leftmargin=*]
    \item \textbf{\textsc{Odfl}:} \textsc{Odfl}~\cite{ODFL} manipulates fine-grained optimization options to construct a single pair of adversarial compilation configurations, namely passing and failing compilation configurations for the same test program. This enables the collection of contrasting coverage spectra of compilers for SBFL-based analysis.
    \item \textbf{DiWi:} DiWi~\cite{chen2019compiler} localizes compiler bugs by applying local operation mutations and employing the Metropolis-Hastings algorithm to sample witness programs.
    \item \textbf{RecBi:} RecBi~\cite{chen2020enhanced} applies structural mutation techniques to generate witness programs and utilizes a reinforcement learning strategy to guide the search for effective witness test programs.
    \item \textbf{LLM4CBI:} LLM4CBI~\cite{LLM4CBI} leverages Large Language Models (LLMs) to generate effective witness programs for compiler bug localization.
    \item \textbf{ETEM:} ETEM~\cite{ETEM} proposes an enhanced program mutation strategy, and further adopts adversarial optimization configurations by \textsc{Odfl} together with improved suspiciousness formulas for computing the fault likelihood of files.
    \item \textbf{\textsc{Basic}:} \textsc{Basic}~\cite{BASIC} identifies the most recent good release and the earliest bad release, and applies a binary search to pinpoint the bug inducing commit (BIC). Files modified in the identified BIC are then regarded as faulty candidates.
\end{itemize}

To the best of our knowledge, \textsc{Odfl} is the first approach that leverages adversarial compilation configurations to localize faults in GCC. In contrast, program mutation-based localization techniques (DiWi, RecBi, LLM4CBI, and ETEM) form a distinct class of approaches that requires substantial time budgets to generate and validate large numbers of mutated programs. Prior studies have demonstrated that \textsc{Odfl} and \textsc{Basic} perform better than program mutation-based techniques in isolating compiler faults~\cite{ODFL,BASIC}. 

Notably, \textsc{Odfl} relies on a single pair of adversarial configurations and performs slightly less than \textsc{Basic} on GCC (20 vs. 21), while being significantly more efficient. In particular, \textsc{Odfl} does not require to installing and testing multiple commit-level compiler versions to identify the BIC. Consequently, although we include all these approaches for completeness, our evaluation emphasizes \textsc{Odfl} as the most directly comparable baseline to demonstrate the effectiveness of \tool's multi-pair of adversarial compilation configuration strategy.

\subsubsection{Measurements}
To assess the effectiveness of the compiler fault localization approaches, we utilize several measurement metrics commonly employed in existing studies~\cite{ODFL, chen2019compiler, chen2020enhanced, LLM4CBI, ETEM}. These metrics allow us to evaluate the performance of the approaches from various perspectives.

\begin{itemize}[leftmargin=*]
    \item \textbf{Top-N:} This metric measures the number of bugs successfully located within the Top-N positions of the ranking list. Specifically, it assesses whether at least one faulty file appears among the Top-N files. Consistent with previous studies~\cite{chen2020enhanced, chen2019compiler, ODFL, LLM4CBI, ETEM}, we consider N values of 1, 5, 10, and 20. A higher value indicates better performance.
    \item \textbf{Mean First Rank (MFR):} MFR calculates the average rank of the first buggy element (i.e., the faulty file in our study) for each bug in the ranking list. This metric focuses on the position of the first identified buggy element, with smaller values indicating better performance.
    \item \textbf{Mean Average Rank (MAR):} MAR calculates the average rank of all buggy elements (i.e., faulty files) for each bug in the ranking list. This metric considers the positions of all buggy elements, with smaller values indicating better performance.
\end{itemize}

\subsection{RQ1: ~\tool~vs. Existing Techniques in Terms of Effectiveness}

\subsubsection{Experimental Settings.}
We adopt the same experimental setup as previous studies~\cite{ODFL, chen2019compiler, chen2020enhanced, LLM4CBI}. Specifically, each witness program-based technique generates mutants continuously until it exceeds one hour of execution time. To mitigate the influence of randomness when replicating these techniques, this study directly utilizes results from prior research that employed the same set of 60 GCC bugs. This ensures a direct and equitable comparison with state-of-the-art techniques.
For a fair comparison, we use the reported results from Tu et al.'s recent study under the default setting (i.e., one-hour termination) as the effectiveness metrics for DiWi, RecBi, and LLM4CBI in our comparison with \tool. For \textsc{Odfl}, ETEM, and \textsc{Basic}, we utilize the best performing results reported in their paper.

\begin{table*}[thp]
  \renewcommand{\arraystretch}{1.5}  %
  \setlength{\tabcolsep}{3.5pt}  %
  \vspace{-1.0em}
  \caption{Compiler fault location effectiveness comparison} 
  \label{tbl:comparison}
  \centering
  \footnotesize
  \def\arraystretch{1.0}
    \vspace{1.0em}   %
    \begin{tabular}{l|cr|cr|cr|cr|cr|cr}
    \toprule
     Technique & Top-1 & $\Uparrow_{Top-1}$ & Top-5 & $\Uparrow_{Top-5}$ & Top-10 & $\Uparrow_{Top-10}$ & Top-20 & $\Uparrow_{Top-20}$ & MFR & $\Uparrow_{MFR}$ & MAR & $\Uparrow_{MAR}$ \\
    \midrule
        \textbf{\tool}      & \textbf{27.00} & —     & \textbf{40.00} & —     & \textbf{51.00} & —     & \textbf{53.00} & —     & \textbf{7.38}  & —     & \textbf{8.53} & —    \\
        \textsc{Odfl}      & 20.00 & 35.00\%  & 31.00 &  29.03\% & 42.00 &  21.43\%  & 47.00 &  12.77\%  & 23.05  & 67.98\% & 23.78 &  64.13\%    \\
         \texttt{DiWi}          & 5.60  & 382.14\% & 19.90 & 101.01\% & 31.30 & 62.94\% & 41.70 & 27.10\% & 22.57 & 67.30\% & 23.10 & 63.07\% \\
         \texttt{RecBi}         & 7.70  & 250.65\% & 23.90 & 67.36\% & 34.20 & 49.12\% & 42.50 & 24.71\% & 20.53 & 64.05\% & 21.06 & 59.50\% \\
         \texttt{LLM4CBI}          & 9.30 & 190.32\% & 24.90 & 60.64\% & 36.40 & 40.11\% & 44.80 & 18.30\% & 15.84 & 53.41\% & 16.35 & 47.83\% \\
         \texttt{ETEM}          & 20.00 & 35.00\% & 38.00 & 5.26\% & 44.00 & 15.91\% & 54.00 &  -1.85\% & 6.63  & -11.31\% & 7.04  &  -21.16\% \\
         \textsc{Basic} & 21.00 & 28.57\% & 36.00 & 11.11\% & 36.00 & 41.37\% & 36.00 & 47.22\% & -- & -- & -- & -- \\
      \bottomrule
        \end{tabular}
        \vspace{-1.5em}
        \caption*{\footnotesize Columns ``$\Uparrow_{*}$'' refers to the improvement rate of~\tool~over a compared technique in terms of the * measurement.}
\vspace{-0.5em}   %
\end{table*}

\subsubsection{Effectiveness.}
The results presented in Table~\ref{tbl:comparison} compare the effectiveness of \tool~with other existing compiler fault localization approaches. The row labeled ``\tool'' shows its performance across various evaluation metrics. Overall, \tool~demonstrates strong bug localization capabilities. Specifically, it successfully locates 27, 40, 51, and 53 compiler bugs within the Top-1, Top-5, Top-10, and Top-20 files of the ranking lists, respectively. This indicates that \tool~effectively isolates approximately 45\%, 67\%, 85\%, and 88\% of the bugs within these categories. Additionally, the Mean First Rank (MFR) and Mean Average Rank (MAR) values for \tool~are 7.38 and 8.53, respectively, suggesting that, on average, the first faulty file appears within the top 7.38 files and all faulty files are located within the top 8.53 files of the ranking list.

When comparing with other approaches (\textsc{Odfl}, DiWi, RecBi, LLM4CBI, ETEM, and \textsc{Basic}) in Table~\ref{tbl:comparison}, we observe that \tool~achieves improved effectiveness in compiler fault localization. Notably, \tool~outperforms the other approaches in the Top-1 and Top-5 metrics, which are particularly valuable for practical applications. Specifically, compared to the state-of-the-art configuration-based approach \textsc{Odfl}, \tool~achieves bug localization for 35.00\%, 29.03\%, 21.43\%, and 12.77\% more bugs within the Top-1, Top-5, Top-10, and Top-20 files, respectively. 
Against the recently proposed ETEM, \tool~isolates 35.00\%, 5.26\%, and 15.91\% more bugs within the Top-1, Top-5, and Top-10 files.
Versus \textsc{Basic}, a common strategy used in practice, \tool~outperforms it by 28.57\%, 11.11\%, 41.37\%, and 47.22\% more bugs within the Top-1, Top-5, Top-10, and Top-20 files.
Furthermore, \tool~achieves the best scores in both MFR and MAR among all baselines except ETEM. These results collectively confirm that \tool~provides more precise and accurate bug localization, establishing a new state of the art over existing approaches.

Despite these improvements, we observe that 7 out of 60 bugs (11.7\%) were not successfully localized within the Top-20 files. We conducted a manual investigation of these cases and found that their localization difficulty stems primarily from fundamental limitations of SBFL techniques, rather than deficiencies specific to \tool. In particular, these challenging bugs exhibit one or more of the following characteristics:

\begin{itemize}[leftmargin=*]
    \item \textbf{Insufficient coverage differentiation}: When passing and failing configurations exhibit highly similar code coverage profiles, SBFL struggles to distinguish faulty files due to minimal differences in execution patterns. 
    \item \textbf{Distributed faults}: Some bugs involve complex interactions across multiple files or modules, resulting in suspiciousness scores being spread thinly across several components. This dilution makes it difficult for SBFL methods to precisely identify the main faulty location.
\end{itemize}

\begin{table}[t]
  \renewcommand{\arraystretch}{1.0}
  \setlength{\tabcolsep}{10pt}
  \caption{Statistic Test Results for \tool~vs State-of-The-Art Techniques in terms of locating GCC bugs}
  \label{tbl:comparisonstatistical}
  \centering
  \small
  \vspace{1.0em}   %
  \begin{tabular}{lcc|lcc|lcc}
    \toprule
    \multicolumn{3}{c|}{\tool~vs. \textsc{Odfl}}
    & \multicolumn{3}{c|}{\tool~vs. DiWi}
    & \multicolumn{3}{c}{\tool~vs. RecBi} \\
    \cmidrule(r){1-3} \cmidrule(lr){4-6} \cmidrule(l){7-9}
    Metric & $p$-value & $\hat{A}_{12}$
    & Metric & $p$-value & $\hat{A}_{12}$
    & Metric & $p$-value & $\hat{A}_{12}$ \\
    \midrule
    Top-1  & $<0.001$ & 1.0 & Top-1  & $<0.001$ & 1.0 & Top-1  & $<0.001$ & 1.0 \\
    Top-5  & $<0.001$ & 1.0 & Top-5  & $<0.001$ & 1.0 & Top-5  & $<0.001$ & 1.0 \\
    Top-10 & $<0.001$ & 1.0 & Top-10 & $<0.001$ & 1.0 & Top-10 & $<0.001$ & 1.0 \\
    Top-20 & $<0.001$ & 1.0 & Top-20 & $<0.001$ & 1.0 & Top-20 & $<0.001$ & 1.0 \\
    MFR    & $<0.001$ & 1.0 & MFR    & $<0.001$ & 1.0 & MFR    & $<0.001$ & 1.0 \\
    MAR    & $<0.001$ & 1.0 & MAR    & $<0.001$ & 1.0 & MAR    & $<0.001$ & 1.0 \\
    \bottomrule
  \end{tabular}
  \vspace{0.5em}
  \begin{tabular}{lcc|lcc|lcc}
    \toprule
    \multicolumn{3}{c|}{\tool~vs. LLM4CBI}
    & \multicolumn{3}{c}{\tool~vs. ETEM}
    & \multicolumn{3}{c}{\tool~vs. \textsc{Basic}} \\
    \cmidrule(r){1-3} \cmidrule(l){4-6}  \cmidrule(l){7-9}
    Metric & $p$-value & $\hat{A}_{12}$
    & Metric & $p$-value & $\hat{A}_{12}$
    & Metric & $p$-value & $\hat{A}_{12}$ \\
    \midrule
    Top-1  & $<0.001$ & 1.0 & Top-1  & $<0.001$ & 1.0 & Top-1  & $<0.001$ & 1.0 \\
    Top-5  & $<0.001$ & 1.0 & Top-5  & $<0.001$ & 1.0 & Top-5  & $<0.001$ & 1.0 \\
    Top-10 & $<0.001$ & 1.0 & Top-10 & $<0.001$ & 1.0 & Top-10 & $<0.001$ & 1.0 \\
    Top-20 & $<0.001$ & 1.0 & \cellcolor{gray!70}Top-20 & \cellcolor{gray!70}$<0.001$ & \cellcolor{gray!70}0.0 & Top-20 & $<0.001$ & 1.0 \\
    MFR    & $<0.001$ & 1.0 & \cellcolor{gray!70}MFR    & \cellcolor{gray!70}$<0.001$ & \cellcolor{gray!70}0.0 & MFR    & — & — \\
    MAR    & $<0.001$ & 1.0 & \cellcolor{gray!70}MAR    & \cellcolor{gray!70}$<0.001$ & \cellcolor{gray!70}0.0 & MAR    & — & — \\
    \bottomrule
  \end{tabular}
\end{table}

\subsubsection{Statistical Test Analysis of Results.} 
In alignment with previous work~\cite{LLM4CBI}, we conduct a Mann-Whitney U-test with a significance level of 0.05 on the total number of bugs located by \tool~versus the state-of-the-art approaches, following the guidelines of Arcuri and Briand~\cite{ArcuriBriand-guide}.
To further mitigate the influence of randomness, we calculate the effect size of the differences between \tool~and the baseline techniques using Vargha and Delaney's $\hat{A}_{12}$ statistics. In this context, given a performance measure \( M \) (e.g., Top-1), the $\hat{A}_{12}$ statistic quantifies the probability that approach A (e.g., \tool) yields higher \( M \) values than approach B (e.g., LLM4CBI). If the two approaches are equivalent, then \( \hat{A}_{12} = 0.5 \). For instance, \( \hat{A}_{12} = 0.9 \) indicates that approach A produces higher results 90\% of the time compared to approach B.

We utilize the R programming language to perform the Mann-Whitney U-test and obtain the p-value. To calculate the effect size, we employ the following formula:

\begin{equation}
    \hat{A}_{12} = \frac{(R1/m) - \left(\frac{m + 1}{2}\right)}{n}
\end{equation}

\noindent where \( R1 \) is the rank sum of the first data group being compared and \( m \) represents the number of observations in the first group and \( n \) represents the number of observations in the second group. The rank sum is a fundamental component of the Mann-Whitney U-test and is provided by most statistical tools. We compute the \( R1 \) value using the function ``$sum(rank(c(X,Y))[seq_along(X)])$'' in R, where \( X \) and \( Y \) are the datasets containing the observations (i.e., metrics over GCC bugs) of the two approaches being compared.

The overall statistical results are presented in Table~\ref{tbl:comparisonstatistical}. A p-value less than 0.05 (often less than 0.001) indicates that \tool~significantly outperforms all existing approaches, including \textsc{Odfl}~\cite{ODFL}, DiWi~\cite{chen2019compiler}, RecBi~\cite{chen2020enhanced}, LLM4CBI~\cite{LLM4CBI}, and ETEM~\cite{ETEM}. Furthermore, we observe that the effect sizes are 1.0 for all techniques except ETEM in the Top-20, MFR, and MAR metrics, suggesting that \tool~has a strong likelihood of yielding better results than existing approaches.

\subsubsection{Analysis of Bugs Uniquely Isolated by \tool.}

To our best knowledge, \textsc{Odfl} is the only prior approach that leverages adversarial compilation configurations to localize faults in GCC compilers. In terms of both localization effectiveness and computational cost, \textsc{Odfl} represents the strongest existing technique for isolating GCC faults. Accordingly, we select \textsc{Odfl} as the primary baseline for direct comparison in our evaluation.

To assess \tool's unique advantages, we further analyze the bugs that can be successfully localized by \tool~but not by \textsc{Odfl}. As shown in Table~\ref{tbl:unibug}, \tool~uniquely localizes 10 bugs that \textsc{Odfl} fails to identify. A closer examination of these cases reveals a consistent pattern: the truly faulty files rarely appear at the very top (Top-1) of any single ranking produced by our three adversarial configuration pairs. Instead, they tend to appear within the Top-5 positions across multiple rankings. The aggregation step in \tool~is therefore critical—it consolidates these consistent, moderate signals from different configuration pairs, elevating the true faulty files above distractor files that only achieve high ranking in a single configuration.

We also observe that \textsc{Odfl}'s effectiveness is comparable to our variant $\mathit{AdvConf}_{2}$, despite their different strategies for producing adversarial configurations. Specifically, $\mathit{AdvConf}_{2}$ constructs a failing configuration by directly disabling half of the bug-free options without any additional search, making it substantially more time-efficient. The additional pairs, $\mathit{AdvConf}_{1}$ and $\mathit{AdvConf}_{3}$, however, significantly improve robustness. These pairs are derived from distinct failing configurations, which increases diversity in both coverage and failure-inducing behaviors, and thereby reducing the likelihood that innocent files are consistently ranked highly. This diversity enables our aggregation step to downweight inconsistently ranked files and to amplify those that are consistently suspicious across multiple configuration pairs.

Concrete examples:

\begin{itemize}[leftmargin=*]
    \item \textbf{GCC Bug \#57719.} The actual faulty file is \texttt{tree-loop-distribution.c}. All other approaches, including \textsc{Odfl}, consistently rank \texttt{lra-assigns.c} as Top-1. In contrast, across our three individual rankings, \texttt{tree-loop-distribution.c} appears at ranks 2, 4, and 4. Meanwhile, \texttt{lra-assigns.c} is ranked Top-1 in two configurations but does not appear in the $\mathit{AdvConf}_{1}$ ranking at all, indicating unstable suspiciousness and suggesting it is an innocent file. Through aggregation, \tool~correctly highlights \texttt{tree-loop-distribution.c} as the most suspicious file, demonstrating how consistent mid-top rankings across diverse configuration pairs can outweigh a single spurious Top-1 result.
    \item \textbf{GCC Bug \#70138 (also discussed in Section 3.2.5).} Both \textsc{Odfl} and $\mathit{AdvConf}_{2}$ incorrectly prioritize \texttt{tree-vect-slp.c}. Under the $\mathit{AdvConf}_{3}$ configuration, however, \texttt{tree-vect-slp.c} drops to rank 67, revealing its instability across configurations. In contrast, the true faulty file, \texttt{tree-vect-loop-manip.c}, remains consistently highly ranked (4, 3, and 4, all within the Top-5). Aggregation enables \tool~to correctly identify the faulty file despite contradictory signals from individual configuration pairs.
\end{itemize}

These findings highlight a fundamental limitation of single-pair localization methods: suspiciousness derived from a single failing–passing pair is vulnerable to incidental coverage overlap. By integrating multiple and diverse configuration pairs, \tool~systematically suppresses inconsistently ranked files and promotes those that repeatedly appear near the top, achieving more accurate and robust fault localization.

\begin{table}[ht]
    \renewcommand{\arraystretch}{0.8}  %
    \setlength{\tabcolsep}{10pt}  %
    \vspace{-0.5em}   %
    \caption{Bugs uniquely isolated by \tool~(\Checkmark: bug can be isolated in Top-1; \XSolidBrush: cannot be isolated)}
    \label{tbl:unibug}
    \centering
    \small
    \vspace{1.0em}   %
    \begin{tabular}{|c|c|c|c|c|c|c|c|}
    \toprule
    \textbf{No.} & \textbf{BugId} & \textbf{\tool} & \textbf{\textsc{Odfl}} & \textbf{No.} & \textbf{Bug ID} & \textbf{\tool} & \textbf{\textsc{Odfl}}\\
    \midrule
1  & 56478 & \Checkmark & \Checkmark  & 31 & 62151 & \Checkmark & \Checkmark \\
2  & 57303 &        &   & 32 & 63551 &        &    \\
3  & 57341 &        &   & 33 & 63605 &        &    \\
4  & 57488 &        &   & 34 & 63659 &        &    \\
5  & \cellcolor{gray!70}57521 & \cellcolor{gray!70}\Checkmark & \cellcolor{gray!70}\XSolidBrush  & 35 & 64041 &        &    \\
6  & \cellcolor{gray!70}57719 & \cellcolor{gray!70}\Checkmark & \cellcolor{gray!70}\XSolidBrush  & 36 & 64682 &        &    \\
7  & 58068 &        &   & 37 & 64756 & \Checkmark & \Checkmark \\
8  & \cellcolor{gray!70}58223 & \cellcolor{gray!70}\Checkmark & \cellcolor{gray!70}\XSolidBrush  & 38 & 64853 &        &    \\
9  & 58343 &        &   & 39 & \cellcolor{gray!70}65318 & \cellcolor{gray!70}\Checkmark &  \cellcolor{gray!70}\XSolidBrush  \\
10 & 58418 &        &   & 40 & 66186 &        &    \\
11 & \cellcolor{gray!70}58451 & \cellcolor{gray!70}\Checkmark & \cellcolor{gray!70}\XSolidBrush  & 41 & 66272 &        &    \\
12 & 58539 &        &   & 42 & 66375 &        &    \\
13 & 58570 &        &   & 43 & 66863 &        &    \\
14 & 58640 &        &   & 44 & 66894 & \Checkmark & \Checkmark \\
15 & 58662 & \XSolidBrush & \Checkmark& 45 & 66952 &        &    \\
16 & 58696 &        &   & 46 & 67121 &        &    \\
17 & 58726 & \XSolidBrush & \Checkmark& 47 & 67456 & \Checkmark & \Checkmark \\
18 & 58759 &        &   & 48 & 67786 & \XSolidBrush & \Checkmark \\
19 & 59221 &        &   & 49 & \cellcolor{gray!70}67828 & \cellcolor{gray!70}\Checkmark &  \cellcolor{gray!70}\XSolidBrush  \\
20 & 59594 &        &   & 50 & 68194 & \Checkmark & \Checkmark  \\
21 & 59715 & \Checkmark & \Checkmark& 51 & 68250 &        &    \\
22 & 59747 & \Checkmark & \Checkmark& 52 & 68990 & \Checkmark & \Checkmark \\
23 & 60452 &        &   & 53 & 69951 & \Checkmark & \Checkmark \\
24 & 61140 & \Checkmark & \Checkmark& 54 & \cellcolor{gray!70}70138 & \cellcolor{gray!70}\Checkmark & \cellcolor{gray!70}\XSolidBrush   \\
25 & 61306 & \Checkmark & \Checkmark & 55 & 70586 & \Checkmark & \Checkmark \\
26 & 61383 & \Checkmark & \Checkmark & 56 & \cellcolor{gray!70}71439 & \cellcolor{gray!70}\Checkmark & \cellcolor{gray!70}\XSolidBrush   \\
27 & 61517 & \Checkmark & \Checkmark& 57 & 71518 &        &    \\
28 & 61518 &        &   & 58 & 80622 & \Checkmark & \Checkmark \\
29 & \cellcolor{gray!70}61578 & \cellcolor{gray!70}\Checkmark & \cellcolor{gray!70}\XSolidBrush  & 59 & 81740 &        &    \\
30 & \cellcolor{gray!70}61681 & \cellcolor{gray!70}\Checkmark & \cellcolor{gray!70}\XSolidBrush  & 60 & 82078 & \Checkmark &  \Checkmark  \\
    \bottomrule
    \end{tabular}
\end{table}

\subsection{RQ2: \tool~vs. Existing Techniques in Terms of Efficiency}

\subsubsection{Experimental Setup.} 
To evaluate the efficiency of \tool, we compared its time consumption with five representative approaches: \textsc{Odfl}, DiWi, RecBi, LLM4CBI, and ETEM.
Our measurements focus exclusively on the core generation phase for each method:
\begin{itemize}[leftmargin=*]
    \item For \textbf{program-based baselines} (DiWi, RecBi, LLM4CBI, and ETEM), we followed the original protocol and used a 1-hour timeout for program mutation. This timeout is configurable, but we adopted the standard setting for consistency.
    \item For \textbf{configuration-based methods} (\tool~and \textsc{Odfl}), we measured the time devoted to generating adversarial compilation configurations.
\end{itemize}
All subsequent procedures—including coverage collection (via \texttt{gcov}) and SBFL analysis (using the Ochiai metric)—were performed identically for all methods and are excluded from our timing data.

\begin{table}[htp]
    \vspace{-1.5em}
    \renewcommand{\arraystretch}{1.0}
    \setlength{\tabcolsep}{18pt}
    \caption{Comparison of Time Consumption in Core Generation Phase (Hours)}
    \vspace{1.0em}
    \label{tbl:efficiency}
    \centering
    \footnotesize
    \begin{tabular}{lccc}
        \toprule
        \textbf{Approach} & \textbf{Average} & \textbf{Maximum} & \textbf{Minimum} \\
        \midrule
        \textbf{\tool}  & 0.0135   & 0.0196 & 0.0103  \\
         \textsc{Odfl}  & 0.0035  & 0.0066  &  0.0019    \\
         DiWi  & 1.0000  & 1.0000  & 1.0000     \\
         RecBi  & 1.0000  & 1.0000  & 1.0000     \\
         LLM4CBI  & 1.0000  & 1.0000  & 1.0000     \\
         ETEM  & 1.0000  & 1.0000  & 1.0000     \\
       \bottomrule
    \end{tabular}
\end{table}

\subsubsection{Experimental Results.} 
Table~\ref{tbl:efficiency} summarizes the results.
As shown, \tool~generates three pairs of adversarial compilation configurations in an average of 0.0135 hours, a runtime that is only marginally higher than \textsc{Odfl}'s 0.0035 hours. By contrast, each program-based baseline (DiWi, RecBi, LLM4CBI, ETEM) consistently reaches the 1-hour timeout for program mutation.

The 1-hour time reported for the baselines reflects only the mutation phase, matching the protocols in their original papers. By excluding post-processing from the timing, we ensure that efficiency comparisons reflect only the core algorithmic steps.

Compared to \textsc{Odfl}, the state-of-the-art configuration-based approach in compiler fault localization, \tool~achieves a substantial efficiency improvement, reducing configuration generation time by nearly 93\%.
This efficiency gain is attributable to \tool's streamlined approach for producing adversarial configurations, in contrast to \textsc{Odfl}'s more complex search strategy designed for maximizing effectiveness per configuration pair.

\subsection{RQ3: Impact of Candidate Filtering on the Effectiveness of~\tool}

\subsubsection{Experimental Setup.} 
Candidate filtering in~\tool~is designed to mitigate the expansive search space inherent to modern compilers such as GCC, which typically comprise hundreds of source files (380 C files on average in our benchmark suite). To rigorously evaluate the impact of candidate filtering on both fault localization effectiveness and computational efficiency, we conduct an ablation study under the following two configurations:

\begin{itemize}[leftmargin=*]
    \item \textbf{\tool}: 
 The complete version of our approach, incorporating a two-stage candidate filtering pipeline:
    \begin{enumerate}
        \item \textit{Execution Trace Filtering}: Excludes source files that are never exercised during compilation, immediately reducing the candidate set by 20\% (from 380 to 304).
        \item \textit{Coverage Differential Filtering}: Further eliminates files whose coverage spectra are identical between passing and failing configurations, yielding an additional 57.5\% reduction (from 304 to 129).
    \end{enumerate}
    \item \textbf{$\mathit{{\tool}_\mathit{wf}}$}: A baseline variant \textbf{w}ithout \textbf{f}iltering, which bypasses both filtering stages and analyzes all 380 files.
\end{itemize}

\noindent Both approaches were evaluated on a consistent set of 60 real-world GCC bugs, ensuring a fair and direct comparison.

\subsubsection{Experimental Results.} 

\begin{enumerate}[leftmargin=*]
    \item \textbf{Improved Fault Localization Accuracy}: As shown in Table~\ref{tbl:ablation study},  enabling candidate filtering increases the Top-1 accuracy of \tool~from 25 to 27 (an 8.00\% improvement) and yields consistent gains across Top-5, Top-10, and Top-20 metrics, as well as ranking-based measures (MFR and MAR).
    \item \textbf{Greater Efficiency}: As shown in Table~\ref{tbl:ablation study efficiency}, candidate filtering reduces the average analysis time by 1.51\% (from 0.0464 to 0.0457 hours). Although the filtering stage introduces a modest overhead, it substantially accelerates the core SBFL computation by reducing the average number of analyzed files from 380 to 129, thereby narrowing the search space and lowering overall computational cost.
\end{enumerate}

\noindent Overall, these results demonstrate that candidate filtering in~\tool~effectively balances search space reduction with localization accuracy, achieving both practical efficiency gains and improved fault localization performance.

\begin{table*}[htp]
  \renewcommand{\arraystretch}{1.5}  %
  \setlength{\tabcolsep}{3.0pt}  %
  \vspace{-1.0em}
  \caption{Comparison of \tool~with and without candidate filtering} 
  \label{tbl:ablation study}
  \centering
  \footnotesize
  \def\arraystretch{1.0}
    \vspace{1.0em}   %
    \begin{tabular}{l|cr|cr|cr|cr|cr|cr}
    \toprule
     Technique & Top-1 & $\Uparrow_{Top-1}$ & Top-5 & $\Uparrow_{Top-5}$ & Top-10 & $\Uparrow_{Top-10}$ & Top-20 & $\Uparrow_{Top-20}$ & MFR & $\Uparrow_{MFR}$ & MAR & $\Uparrow_{MAR}$ \\
    \midrule
        \textbf{\tool}      & \textbf{27} & —     & \textbf{40} & —     & \textbf{51} & —     & \textbf{53} & —     & \textbf{7.38}  & —     & \textbf{8.53} & —    \\
    \midrule
        $\mathit{{\tool}_\mathit{wf}}$      & 25 & 8.00\%     & 38 & 5.26\%     & 49 & 4.08\%    & 51 & 3.92\%     & 9.52  & 22.48\%     & 10.96 & 22.17\%    \\
    \bottomrule
        \end{tabular}
        \vspace{-1.0em}
        \caption*{\footnotesize Columns ``$\Uparrow_{*}$'' refers to the improvement rate of~\tool~over $\mathit{{\tool}_\mathit{wf}}$ in terms of the * measurement.}
\vspace{-0.5em}   %
\end{table*}

\begin{table*}[htp]
    \renewcommand{\arraystretch}{1.0}
    \setlength{\tabcolsep}{8pt}
    \caption{Compiler fault location efficiency comparison (hours)}
    \vspace{1.0em}   %
    \label{tbl:ablation study efficiency}
    \centering
    \footnotesize
    \begin{tabular}{l|cc|cc|cc}
        \toprule
        \textbf{Approach} & \textbf{Average} & $\Downarrow$ & \textbf{Maximum} & $\Downarrow$ & \textbf{Minimum} & $\Downarrow$ \\
        \midrule
        \textbf{\tool} & \textbf{0.0457} & 1.51\% & \textbf{0.1014} & 9.22\% & 0.0178 & -6.59\% \\
        $\mathit{{\tool}_\mathit{wf}}$ & 0.0464 & — & 0.1117 & — & \textbf{0.0167} & — \\
        \bottomrule
    \end{tabular}
    \vspace{-1.0em}
    \caption*{\footnotesize Columns ``$\Downarrow$'' refer to the reduction rate of~\tool~over $\mathit{{\tool}_\mathit{wf}}$ in time consumption.}
\end{table*}

\subsection{RQ4: Impact of the Number of Adversarial Compilation Configuration Pairs on \tool's Effectiveness}

\subsubsection{Experimental Setup.} 
To investigate the effect of the number of adversarial configuration pairs, we systematically varied the number of pairs used by \tool~(i.e., 1, 3, 5, and 10) and analyzed the resulting trade-offs in accuracy and efficiency.
For the single-pair scenario, we introduced three variants, each corresponding to a different adversarial configuration pair and omitting the rank aggregation step, denoted as $\mathit{AdvConf}_{1}$, $\mathit{AdvConf}_{2}$, and $\mathit{AdvConf}_{3}$ (detailed in Section 3.2.3). This design allows us to isolate the benefits of rank aggregation in \tool. Table~\ref{tbl:Pair Numbers} presents a comparative analysis of these variants and \tool. We further examine the distribution of correctly localized bugs using the Top-1 and Top-5 metrics.
For experiments with three adversarial pairs, the full \tool~algorithm is employed. The three configurations respectively disable all, half, and none of the bug-free options. For five pairs, we add two more configurations that disable a quarter and three-quarters of the bug-free options. In the ten-pair setting, configurations are distributed more finely: one disables all, one disables none, and the remaining eight disable increments from one-eighth to seven-eighths of the bug-free options. This setup enables a comprehensive assessment of \tool~across a diverse spectrum of adversarial scenarios. The results, including localization effectiveness and the impact of rank aggregation, are summarized in Table~\ref{tbl:Pair Numbers}.
Efficiency results, in terms of average, maximum, and minimum running time, are provided in Table~\ref{tbl:Pair Numbers efficiency}.

\begin{table*}[thp]
  \renewcommand{\arraystretch}{1.5}  %
  \setlength{\tabcolsep}{3.0pt}  %
  \vspace{-1.0em}
  \caption{Trade-offs of Varying Pair Numbers} 
  \label{tbl:Pair Numbers}
  \centering
  \footnotesize
  \def\arraystretch{1.0}
    \vspace{1.0em}   %
    \begin{tabular}{l|cr|cr|cr|cr|cr|cr}
    \toprule
     Technique & Top-1 & $\Uparrow_{Top-1}$ & Top-5 & $\Uparrow_{Top-5}$ & Top-10 & $\Uparrow_{Top-10}$ & Top-20 & $\Uparrow_{Top-20}$ & MFR & $\Uparrow_{MFR}$ & MAR & $\Uparrow_{MAR}$ \\
     \midrule
        \texttt{$\mathit{AdvConf}_{1}$} & 18 & 50.00\% & 34 & 17.65\% & 44  & 15.91\% & 51 & 3.92\%  & 8.61 & 14.29\% & 9.92 & 14.01\%  \\
	\texttt{$\mathit{AdvConf}_{2}$}   & 16 & 68.75\% & 33 & 21.21\% & 46 & 10.87\% & 54 & -1.85\% & 8.73 & 15.46\% & 9.54 & 10.59\% \\
	\texttt{$\mathit{AdvConf}_{3}$} & 10 & 170.00\% & 35 & 14.29\% & 46 & 10.87\% & 52 & 1.92\% & 9.78 & 24.54\% & 10.71 & 20.35\% \\
     \midrule
        \textbf{\tool}      & \textbf{27} & —     & \textbf{40} & —     & \textbf{51} & —     & 53 & —     & 7.38  & —     & 8.53 & —    \\
    \midrule
        ${\tool}_\mathit{5p}$      & 25 & 8.00\%     & \textbf{40} & 0.00\%     & 46 & 10.87\%    & 54 & -1.85\%     & \textbf{7.25}  & -1.79\%     & \textbf{8.22} & -3.77\%    \\
        ${\tool}_\mathit{10p}$      & 24 & 12.50\%     & 39 & 2.56\%     & \textbf{51} & 0.00\%    & \textbf{55} & -3.64\%     & 7.37  & -0.01\%     & 8.52 & -0.01\%    \\
      \bottomrule
        \end{tabular}
        \vspace{-1.0em}
        \caption*{\footnotesize Columns ``$\Uparrow_{*}$'' refers to the improvement rate of~\tool~over other variants in terms of the * measurement.}
\vspace{-0.5em}   %
\end{table*}

\begin{table}[thp]
    \vspace{-0.5em}
    \renewcommand{\arraystretch}{1.0}
    \setlength{\tabcolsep}{8pt}
    \caption{Different Pair Numbers efficiency comparison (hours)}
    \vspace{1.0em}
    \label{tbl:Pair Numbers efficiency}
    \centering
    \footnotesize
    \begin{tabular}{lccc}
        \toprule
        \textbf{Approach} & \textbf{Average} & \textbf{Maximum} & \textbf{Minimum} \\
        \midrule
        ${\tool}_{1p}$  & 0.0214  & 0.0412  & 0.0079     \\
        \textbf{\tool}  & 0.0457 & 0.1014 & 0.0178  \\
         ${\tool}_{5p}$  & 0.0689 & 0.1601  &  0.0244    \\
         ${\tool}_{10p}$  & 0.1286  & 0.3064  &  0.0350   \\
       \bottomrule
    \end{tabular}
    \vspace{-0.5em}
\end{table}

\begin{figure}[th]
    \centering
    \begin{subfigure}[t]{0.48\textwidth}
        \centering
        \includegraphics[width=\textwidth]{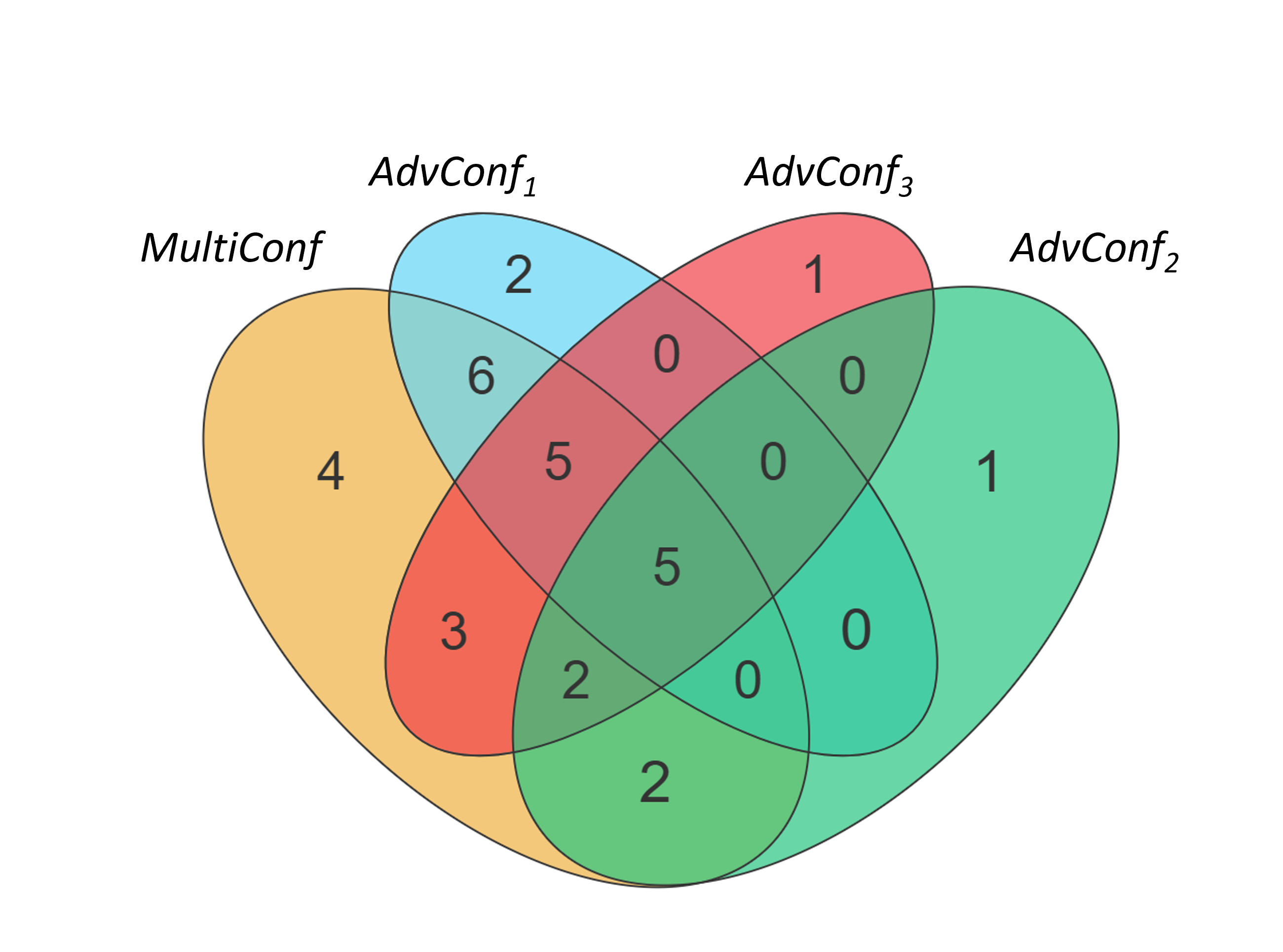}
        \caption{Top-1}\label{fig:venn-compare1}
    \end{subfigure}
    \hspace{0.01\textwidth} %
    \begin{subfigure}[t]{0.48\textwidth}
        \centering
        \includegraphics[width=\textwidth]{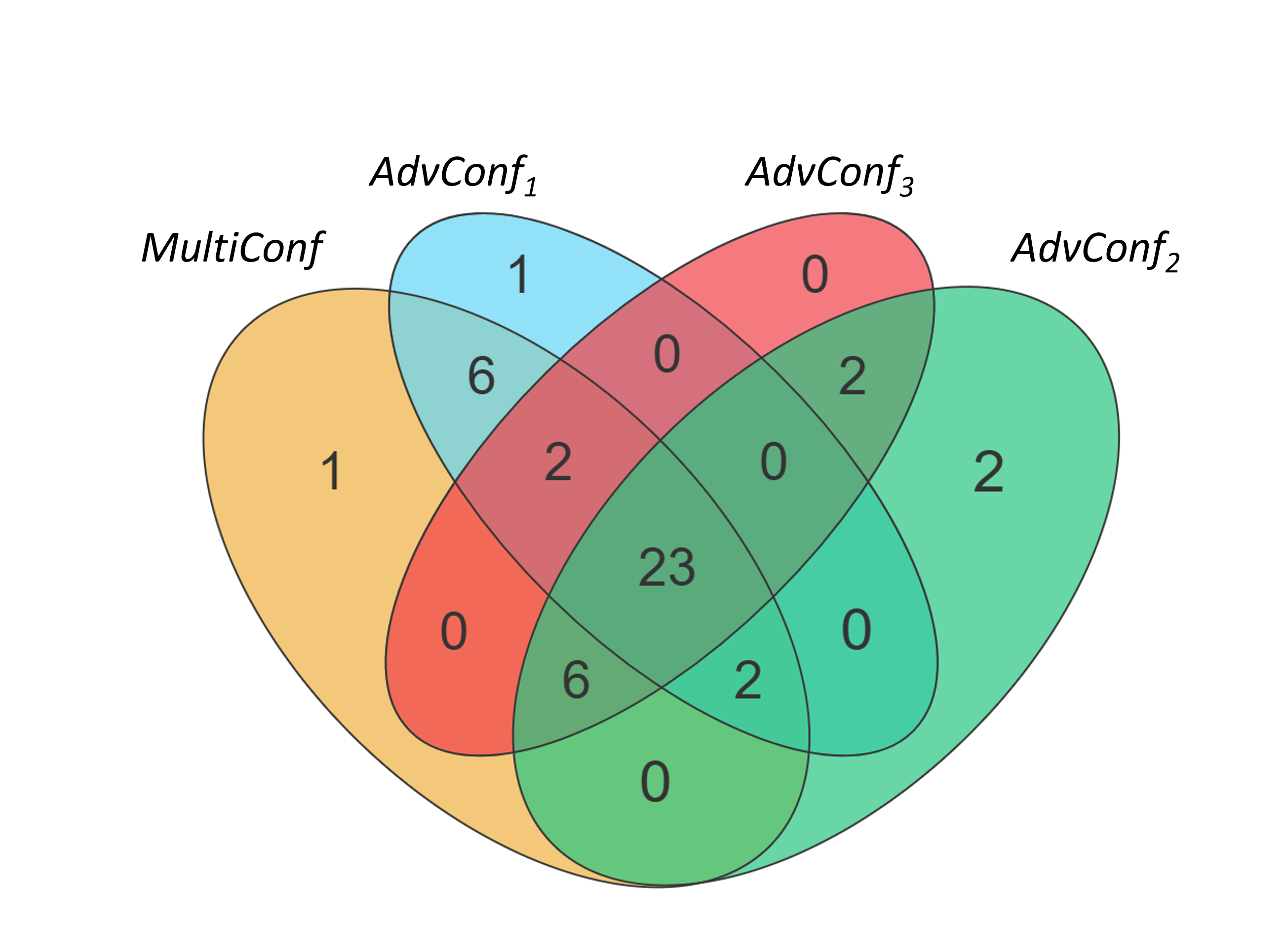}
        \caption{Top-5}\label{fig:venn-compare2}
    \end{subfigure}
    \caption{Distribution of correct localization achieved by \tool~and individual pairs (Top-1 and Top-5).} \label{fig:venn}
\end{figure}

\subsubsection{Experimental Results.} 
Our results show that different adversarial pairs have varying abilities to localize compiler bugs, supporting the rationale for using rank aggregation in \tool. As illustrated in Figure~\ref{fig:venn}, some bugs are uniquely localized by \tool~when aggregating across pairs, while many bugs identified by individual pairs are also captured by \tool.

Table~\ref{tbl:Pair Numbers} demonstrates that \tool~consistently outperforms single-pair variants, especially in the Top-1 metric, achieving improvements of 50\%, 68.75\%, and 170\% over $\mathit{AdvConf}_{1}$, $\mathit{AdvConf}_{2}$, and $\mathit{AdvConf}_{3}$, respectively. This highlights the effectiveness of rank aggregation in combining the strengths of multiple adversarial configurations for more accurate bug localization.

Interestingly, Figure~\ref{fig:venn-compare1} shows that 4 out of 27 bugs localized by \tool~in the Top-1 setting are not found by any individual pair, illustrating the complementary power of aggregation. Similar patterns are observed for Top-5 in Figure~\ref{fig:venn-compare2}.

We further investigate the impact of increasing the number of adversarial configuration pairs. As shown in Tables~\ref{tbl:Pair Numbers} and~\ref{tbl:Pair Numbers efficiency}, our results reveal that three pairs offer the best balance between accuracy and efficiency:
\begin{itemize}[leftmargin=*]
    \item \textbf{Saturation Effect}: Accuracy plateaus at three pairs (27 Top-1 in Table~\ref{tbl:Pair Numbers}), with marginal or no gains beyond this point. Specifically, while increasing from one pair to three pairs brings significant improvements across all metrics\textemdash for example, Top-1 rises from 18 to 27 (a 50\% increase), and MFR drops from 8.61 to 7.38\textemdash using more than three pairs does not yield further substantial gains. When five pairs are used, Top-1 actually drops to 25, and Top-5 remains at 40, identical to the three-pair configuration. With ten pairs, Top-1 further decreases to 24, and gains in other metrics are either negligible or negative.
    \item \textbf{Efficiency Trade-off}: Increasing the number of adversarial pairs has a notable impact on computational efficiency. The runtime for the three-pair configuration is 0.0457 hours on average, but this increases threefold to 0.1286 hours when using ten pairs, as shown in Table~\ref{tbl:Pair Numbers efficiency}. The computational overhead thus scales much faster than accuracy improvements, leading to an inefficient use of resources at higher pair counts.
\end{itemize}

\noindent While generating more pairs is technically feasible, our empirical evidence indicates that using three adversarial pairs achieves near-optimal accuracy with minimal computational overhead. Increasing the number further yields diminishing returns and higher costs, making three pairs the recommended configuration for practical use.

\subsection{RQ5: Effect of Different SBFL Formulae in~\tool}

\begin{table}[t]
    \vspace{-1.0em}
	\centering
	\caption{\small The six adapted SBFL formulae in \tool}
	\label{formula}
	\small
	\setlength\tabcolsep{15pt} %
	\def\arraystretch{1.8} %
    \vspace{1.0em}   %
	\begin{tabular}{cc}
		\toprule
		Name & Formula \\ \hline
		\textbf{Ochiai} \cite{abreu2007accuracy} & $\displaystyle \frac{ef_{s}}{\sqrt{(ef_{s}+nf_{s})(ef_{s}+ep_{s})}}$ \\[1em]
		\textbf{Tarantula} \cite{jones2005empirical} & $\displaystyle \frac{ef_{s}/totalf_s}{ef_s/totalf_s + ep_s/totalp_s}$ \\[1em]
		\textbf{DStar} \cite{wong2013dstar} & $\displaystyle \frac{ef_{s}^\star}{ep_{s} + nf_{s}}$ \\[1em]
		\textbf{Dice} \cite{wong2016survey} & $\displaystyle \frac{2ef_{s}}{ef_{s} + nf_{s} + ep_{s}}$ \\[1em]
		\textbf{Barinel} \cite{abreu2009spectrum} & $1 - \frac{ep_{s}}{ep_{s} + ef_{s}}$ \\[1em]
		\textbf{Op2} \cite{naish2011model} & $ef_{s} - \frac{ep_{s}}{(totalp_{s} + 1)}$ \\ 
		\bottomrule
		\multicolumn{2}{l}{\scriptsize We used $\star = 2$, the most thoroughly explored value \cite{pearson2017evaluating}}
	\end{tabular}
\vspace{-0.5em}   %
\end{table}

\subsubsection{Experimental Setup.} 
\tool~defaults to using the Ochiai formula~\cite{abreu2007accuracy} to calculate the suspiciousness value of statements in the compiler. This choice is based on previous reports that have identified Ochiai as the most effective formula for SBFL \cite{wong2016survey, pearson2017evaluating}. However, it remains unknown whether \tool~can achieve similar performance using alternative SBFL formulae.
To explore the impact of different formula choices on the performance of \tool, we compare Ochiai with five other most widely studied SBFL formulae~\cite{wen2019historical, wong2016survey, pearson2017evaluating}. This comparative analysis aims to offer insights into the effectiveness of various formulae within the context of \tool~and its ability to accurately localize compiler bugs.
An overview of these SBFL formulae is provided in Table~\ref{formula}.
In these formulae, $ef_s$, $nf_s$, and $ep_s$ are defined in the same way as explained in Formula~\ref{formula:ochia}, $np_s$ denotes the number of passing cases that do not execute the statement $s$, and $totalf_s$ and $totalp_s$ denote the total number of failing cases and passing cases for statement $s$ respectively.
We studied how different formulae affect the performance of \tool~and evaluated them with Top-1, Top-5, Top-10, Top-20, MAR and MFR.

\subsubsection{Experimental Results.} 
Table~\ref{tbl:formulae} presents the results of \tool~using different formulae.
In this table, different rows represent the different SBFL formulae utilized and different columns represent the different metrics evaluated.
It can be observed that \tool~achieves almost similar results when adopting the formulae Ochiai, Tarantula, Dstar, Dice and Barinel.
This explains why most of the formulae have no obvious effect on the performance of \tool~and reveals the effectiveness and stability of \tool.
However, all these formulae are not specifically designed for locating compiler bugs, and whether more effective formulae could be utilized remains unknown. We leave such exploration as our future work.

\begin{table}
    \renewcommand{\arraystretch}{1.2}
    \setlength{\tabcolsep}{5pt}
    \vspace{-1.0em}
    \caption{Performance of~\tool~using different SBFL formulae}
    \vspace{1.0em}
    \label{tbl:formulae}
    \centering
    \footnotesize
    \begin{tabular}{lcccccc}
        \toprule
        \textbf{Formula} & \textbf{Top-1} & \textbf{Top-5} & \textbf{Top-10}  & \textbf{Top-20} & \textbf{MAR}  & \textbf{MFR} \\
        \midrule
        Ochiai & \textbf{27}  & 40   & \textbf{51} & \textbf{53}  & 7.38 & 8.53 \\
        Tarantula & 22 & \textbf{41} & 50 & \textbf{53}  & 7.73 & 8.81 \\
        DStar    & 22 & 38 & 44 & 48 & 10.77 & 11.78 \\
        Dice  & 25 & \textbf{41}   & 48 & \textbf{53}  & \textbf{7.28} & \textbf{8.30} \\
        Barinel  & 25 & \textbf{41} & 48 & \textbf{53} & \textbf{7.28} & 8.31 \\
        Op2 & 22 & 38 & 49 & 52  & 8.27 & 9.23 \\
        \bottomrule
    \end{tabular}
\end{table}

\section{Discussion}
\label{sec:validity}
\subsection{Limitations and Scope}

\tool~is specifically designed to localize the faulty files associated with option-sensitive compiler bugs—those whose manifestation is influenced by the selection of compiler options, particularly optimization flags. According to the official GCC bug taxonomy\footnote{\url{https://gcc.gnu.org/bugzilla/describekeywords.cgi}}, our approach is most applicable to the following primary categories:

\begin{itemize}[leftmargin=*]
    \item \textbf{Wrong code}: Bugs where the compiler produces incorrect output code. These issues frequently depend on particular optimization or configuration settings. By analyzing behavioral differences under various options, our technique can help pinpoint the source files most likely responsible for such failures.
    \item \textbf{Ice on valid code}: Internal compiler errors (ICEs) or assertion failures encountered when compiling valid source code. These are often triggered by specific combinations of compiler flags, and our approach is effective in localizing the files at fault in these scenarios.
\end{itemize}

\noindent There are several limitations to our technique:

\begin{itemize}[leftmargin=*]
    \item \textbf{Option independence:} \tool~is not intended for bugs whose manifestation is independent of compiler options; such bugs fall outside the scope of our localization strategy.
    \item \textbf{Option Coverage}: Accuracy may be reduced for bugs triggered only by rare or highly specific option combinations not included in our experimental setup.
    \item \textbf{Compiler Specificity}: Our evaluation focuses on GCC. Adapting the method to compilers with different optimization infrastructures (e.g., LLVM) may require additional work.
\end{itemize}

\subsection{Witness Programs from Other Compilers}

Recent work has shown that real-world programs from bug reports (even from other compilers), as opposed to artificially constructed or randomly generated tests, can significantly enrich compiler testing~\cite{zhong2023test}. These real-world programs are typically more representative of practical usage scenarios and often trigger subtler, harder-to-find compiler defects. They can also be treated as witness programs in fault localization scenario.

However, incorporating real-world witness programs into a localization framework presents notable challenges. The effectiveness of spectrum-based fault localization relies on discriminative coverage information. Existing witness-based approaches typically minimize code-coverage differences between bug-triggering and witness programs with only minor variations, thereby facilitating the identification of suspicious compiler modules responsible for a fault. If additional witness programs are not carefully selected, especially when sourced from other compilers or unrelated bug reports, the resulting coverage data may become noisy, making it harder to distinguish faulty components from non-faulty ones. This risk is especially pronounced in large, complex compiler codebases where even minor differences in exercised code paths can significantly affect localization outcomes.

Therefore, while integrating real-world seeds holds promise for expanding test coverage and bug detection, their indiscriminate use in fault localization must be approached with caution. A promising direction for future work is the development of hybrid strategies that combine the strengths of both targeted witness programs and adversarial configuration pairs, balancing increased robustness with precise localization.

\subsection{Relation to Program Reduction}

Program reduction techniques, such as hierarchical delta debugging (HDD)~\cite{Misherghi2006}, probabilistic delta debugging (PDD)~\cite{Wang2021}, cause reduction~\cite{Groce2016}, and C-Reduce~\cite{Regehr2012}, play a pivotal role in the testing and debugging of compilers and other complex software systems~\cite{10.1145/3763093}. Their primary objective is to shrink complex, failure-inducing test cases to minimal programs that still trigger the same compiler bug. This minimization improves human readability, facilitates bug reproduction, and streamlines manual analysis.

These reduction methods directly complement compiler fault localization techniques like \tool. Once a test case has been reduced, the compiler is exercised on a smaller, more focused input, causing only a subset of its codebase to be executed. This has two main benefits for localization: (1) it eliminates noise by excluding compiler code unrelated to the bug, and (2) it increases the precision of coverage-based analyses, as the remaining execution paths are more likely to be causally linked to the failure. As a result, \tool~can leverage these cleaner, more targeted traces to pinpoint the underlying faulty compiler components more accurately and efficiently.

While program reduction and fault localization address different stages of the debugging workflow, they are highly synergistic: reduction enhances the effectiveness of localization, but does not replace the need for specialized techniques to map failures in test programs to faulty source code within the compiler.

\subsection{Threats to Validity}

\subsubsection{Internal Validity}

\begin{enumerate}[leftmargin=*]
    \item \textbf{Configuration Selection Bias}: Although our systematic approach to generating adversarial configurations (Section 3.2.3) reduces arbitrary selection, results may still be influenced by the specific configurations chosen. We address this by utilizing multiple configuration pairs to diversify coverage and by applying a weighted aggregation mechanism to further limit dependence on any single configuration pair.
    \item \textbf{SBFL Formula Sensitivity}: As shown in our experiments (RQ5), Top-1 accuracy varies by less than 5\% across six SBFL formulae (Ochiai, Tarantula, etc.), suggesting our method is robust to formula selection. We select Ochiai as the default due to its consistently strong performance.
    \item \textbf{Impact of Filtering}: The candidate filtering step reduces the search space from 380 to 129 files, improving efficiency. However, in rare cases where coverage differences are subtle, this process may inadvertently exclude relevant files. While such cases did not occur in our benchmark, further investigation is needed to assess their frequency and impact in broader settings.
    \item \textbf{Randomness in Baseline Comparisons}: Some baseline approaches (e.g., DiWi, RecBi, LLM4CBI, ETEM) involve randomized mutations, introducing variance in fault localization outcomes. To control for this, we report results from prior controlled experiments, minimizing the influence of randomness on comparative analysis.
\end{enumerate}

\subsubsection{External Validity}
\begin{enumerate}[leftmargin=*]
    \item \textbf{Compiler Specificity}: Our approach is tailored to GCC’s optimization infrastructure and does not directly generalize to other ordered-pass-based compilers like LLVM (see Section 4.1.1). Adapting our methodology to different architectures remains future work.
    \item \textbf{Benchmark Size}: While our evaluation on 60 bugs matches the scale of prior work, this sample may not fully capture the diversity of real-world scenarios. Expanding our benchmark set and including a broader range of bug types are important for strengthening external validity.
\end{enumerate}

\subsection{Future Work}
Several promising directions remain for extending this research:
\begin{enumerate}[leftmargin=*]
    \item \textbf{Minimal Trigger Set Identification}: At present, when no individual bug-triggering options can be isolated, we generate adversarial configuration pairs at the optimization level granularity (see Section 3.2.3). However, to more effectively address complex interactions among optimization options, a promising direction for future work is to identify minimal sets of options that collectively cause the bug. This approach could yield more precise insights than operating solely at the optimization level granularity.
    \item \textbf{Hybrid with Witnesses}: \tool~is orthogonal to witness-program-based approaches. Combining a carefully selected set of real or witness programs with multiple adversarial configuration pairs may further enhance the robustness and effectiveness of fault localization.
    \item \textbf{Live Bug Evaluation}: Beyond benchmarked bugs, applying \tool~to current, unresolved compiler issues and engaging with developers to obtain feedback would provide stronger evidence for the practical utility and real-world impact of our method.
    \item \textbf{Benchmark Expansion}: To improve external validity, we intend to expand our benchmark suite, including more bugs and a wider variety of bug types for a more comprehensive evaluation.
    \item \textbf{Cross-Compiler Generalization}: We intend to adapt our approach to other compilers, such as LLVM, by evolving our methodology to accommodate architectural differences in optimization pipelines and pass management.
\end{enumerate}

These future directions aim to address current limitations, strengthen the validity of our results, and broaden the impact of our approach in real-world compiler development and maintenance.

\section{Related Work}
\label{sec:related-work}
Software fault localization has been extensively studied, resulting in a rich body of literature encompassing a wide range of techniques. Representative approaches include spectrum-based fault localization (SBFL)~\cite{abreu2009spectrum, tang2017accuracy, wen2019historical}, slicing-based techniques~\cite{agrawal1993debugging, zhang2005experimental}, mutation-based approaches~\cite{moon2014ask, hong2015mutation, papadakis2015metallaxis}, and model-based fault localization~\cite{jung2018combining, wen2016locus}. Among these, SBFL techniques~\cite{abreu2009spectrum, tang2017accuracy} have received the most research attention, as they quantify the suspiciousness of program elements by contrasting their execution coverage in failing and passing test cases.

Compiler fault localization introduces additional challenges due to the complexity of compilers and the difficulty of constructing suitable test cases. Witness program-based approaches, such as DiWi~\cite{chen2019compiler}, RecBi~\cite{chen2020enhanced}, LLM4CBI~\cite{LLM4CBI}, and ETEM~\cite{ETEM}, attempt to address challenges by generating both passing and failing test programs to isolate compiler faults. These approaches leverage SBFL to rank suspicious compiler files. However, they often suffer from instability in test generation, high computational costs, and limited diversity in the generated witness programs.

\textsc{Odfl}~\cite{ODFL} proposed adversarial compilation configurations as an alternative to witness program generation. By selectively disabling fine-grained compiler optimization options, \textsc{Odfl} constructs passing and failing configurations for SBFL-based analysis. Although effective, \textsc{Odfl} relies on a single adversarial configuration pair, which constrains coverage diversity and leads to suboptimal localization performance.

\textsc{Basic}~\cite{BASIC} revisited compiler fault localization from the perspective of practical debugging workflows by directly identifying the bug-inducing commit (BIC) and treating the files modified in that commit as faulty candidates. Its evaluation demonstrated that \textsc{Basic} achieves performance comparable to, and in many cases better than, state-of-the-art SBFL-based compiler fault localization techniques. Nevertheless, \textsc{Basic} incurs substantial computational overhead, as it requires installing and testing multiple commit-level compiler versions to identify the BIC. 

To address these limitations, we propose \tool, a novel approach that generates multiple pairs of adversarial compilation configurations. By aggregating rankings from these configuration pairs, \tool~substantially increases coverage diversity and improves localization accuracy. This design is inspired by ensemble learning, in which combining diverse predictors mitigates individual bias~\cite{ZHOU2002239,zhou2025ensemble}. 
Our empirical evaluation on real-world GCC bugs demonstrates that \tool~consistently outperform existing compiler fault isolation techniques.

\section{Conclusion}
\label{sec:conclusion}
In this study, we introduce \tool, a novel approach for isolating compiler faults by leveraging multiple pairs of adversarial compilation configurations. By aggregating ranks from diverse configuration pairs, \tool~represents a significant advancement in fault localization accuracy.
Our experimental results demonstrate the substantial improvements of \tool~over existing approaches in both effectiveness and efficiency. Specifically, \tool~successfully localizes 27 out of the 60 GCC bugs within the Top-1 file, which is critically important for practical applications, reflecting an impressive 35.0\% improvement over \textsc{Odfl}. 
The artifact of this work is publicly available at: \textbf{\url{https://zenodo.org/records/18051063}}.

\section{ACKNOWLEDGMENTS}

We are grateful to the anonymous reviewers and the editorial team for their valuable and constructive feedback. We would like to express our appreciation to the authors of DiWi, RecBi, LLM4CBI, \textsc{Odfl}, and \textsc{Basic} for generously sharing their tools and data. This work was supported in part by the National Natural Science Foundation of China (Grants 624B2067, 62572226, and 62472215), the Jiangsu Natural Science Foundation (Grant BK20231402), the Collaborative Innovation Center of Novel Software Technology and Industrialization.

\bibliographystyle{ACM-Reference-Format}
\normalem
\bibliography{full}

\end{document}